\documentclass[11pt]{article}
\usepackage{fullpage}
\usepackage{amsmath,amssymb}
\usepackage{amsthm}
\usepackage{bm}
\usepackage{subfigure,graphicx}
\usepackage{cite}
\usepackage{url}
\usepackage{setspace}

\newcommand{\btheta}{\mbox{\boldmath $\theta$}}

\newcommand{\bpi}{\mbox{\boldmath $\pi$}}
\newcommand{\btau}{\mbox{\boldmath $\tau$}}

\newcommand{\Amat}{{\bf A}}
\newcommand{\bx}{{\bf x}}
\newcommand{\bw}{{\bf w}}
\newcommand{\br}{{\bf r}}
\newcommand{\by}{{\bf y}}
\newcommand{\1}[1]{\mathbb{I}_{\{#1\}}}
\newcommand{\bp}{\mathbf{p}}

\newcommand{\bz}{{\bf z}}

\newcommand{\thetamin}{\theta_{\min}}
\newcommand{\E}{\mathbb{E}}

\newcommand{\var}{\mbox{Var}}

\newtheorem{thm}{Theorem}
\newtheorem{prop}{Proposition}

\newtheorem{cor}{Corollary}

\begin{document}

\title{\bf Network Inference from Co-Occurrences}
\author{Michael G. Rabbat, M\'{a}rio A. T. Figueiredo, and Robert D.
Nowak} %

%

\maketitle


\renewcommand{\thefootnote}{}
\footnotetext{\hspace{-2\parindent} \newline\indent M.G.~Rabbat
and R.D.~Nowak are with the Department of Electrical and Computer
Engineering, University of Wisconsin, Madison, WI, 53706. Email:
\texttt{rabbat@cae.wisc.edu, nowak@engr.wisc.edu}. \newline\indent
M.A.T.~Figueiredo is with {\it Instituto de
Telecomunica\c{c}\~{o}es} and the Department of Electrical and
Computer Engineering, {\it Instituto Superior T\'ecnico}, Lisboa,
Portugal. Email: \texttt{mario.figueiredo@lx.it.pt}.}
\renewcommand{\thefootnote}{\arabic{footnote}}

\doublespacing

\begin{abstract}
  The discovery of network structures is a fundamental problem in
  arising in numerous fields of science and technology, communication
  systems, biology, sociology and neuroscience.  Unfortunately, it is
  often difficult to obtain data that directly reveals network
  structure, and so one must infer a network from incomplete data.
  This paper considers inferring network structure from
  ``co-occurrence'' data; observations that identify which network
  components (e.g., switches, routers, genes) carry each transmission
  but does not indicate the order in which they handle the
  transmissions.  Without order information, there is an exponential
  number of feasible networks that are compatible with the observed
  data. Yet, the basic physical principles underlying most networks
  strongly suggest that all feasible networks are not equally likely.
  In particular, network elements that co-occur in many observations
  are probably closely connected.  We model the co-occurrence
  observations as independent realizations of a random walk on the
  underlying graph, subjected to a random permutation which accounts
  for the lack of order information.  Treating the permutations as
  missing data, we derive an exact \emph{expectation-maximization}
  (EM) algorithm for estimating the random walk parameters.  The model
  and EM algorithm significantly simplify the problem, but the
  computational complexity of the reconstruction process does grow
  exponentially in the length of the longest transmission path.  For
  large networks the exact E-step may be computationally intractable,
  and so we also propose an efficient \emph{Monte Carlo EM} (MCEM)
  algorithm, based on importance sampling, and derive conditions which
  ensure convergence of the algorithm with high probability.
  Remarkably, the MCEM maintains the desirable properties of the exact
  EM algorithm and reduces the complexity of each iteration to
  polynomial-time.  Simulations and experiments with Internet
  measurements demonstrate the promise of this approach.
\end{abstract}


\section{Network Inference and Co-Occurrence Observations}
\label{introduction}
The study of complex networked systems is an emerging field,
impacting nearly every area of engineering and science, including
the important domains of communication systems, biology,
sociology, and cognitive science.  The analysis of communication
networks enables a better understanding of routing, transmission
patterns, and information flow \cite{rabbat05,coates02}. The
analysis of biological networks provides insight into the
functional roles played by different genes, proteins, and
metabolites in biological systems \cite{klipp05,palsson06}. The
analysis of social networks contributes to a deeper understanding
of interactions, dynamics, and the structure of organizations
\cite{wasserman94,newman06}. The analysis of functional
connectivity networks in the brain is necessary for the
understanding of couplings and interactions between different
neuronal colonies \cite{sporns02,sporns04,brainConnectivity05}.
Obtaining or inferring the structure of networks from experimental
data precedes any such analysis and is thus a basic and
fundamental task, critical to many applications.

Unfortunately, measurements which directly reveal network structure
are often beyond experimental capabilities or are excessively
expensive.  This paper considers inferring network structure from
observations that identify which network components (e.g., switches,
routers, genes) carry each transmission but does not indicate the
order in which they handle the transmissions.  Mathematically, the
underlying network structure can be represented as a directed graph,
and the vertices involved in each transmission form a connected
subgraph.  The observations only reflect which subset of vertices are
involved or ``occur'' in each transmission; not their
inter-connectivity.  We refer to such observations as
\emph{co-occurrences}.  Co-occurrence observations arise naturally in
each of the application areas mentioned above.

Transmissions over telecommunication networks are carried by links and
routers/switches which form a path between the source and terminal
nodes.  In some cases, it is impossible to directly observe the order
in which the routers/switches handle each transmission, since sensors
are geographically distributed, making precise time-synchronization
impractical.  The so-called \emph{internally-sensed network
  tomography} problem specifically aims at recovering network
structure from unordered lists of network elements along transmission
paths \cite{rabbat05}.

Biological signal transduction networks describe fundamental cell functions
such as growth, metabolism, differentiation, and apoptosis
(disintegration) \cite{palsson06}.  Although it is possible to
test for individual, localized interactions between genes pairs,
such experiments are expensive and time-consuming. High-throughput
measurement techniques such as microarrays have successfully been
used to identify the components of different signal transduction
pathways \cite{zhu05}.  However, microarray data only reflects
order information at a very coarse, unreliable level.  Developing
computational techniques for inferring pathway orders is an active
research area \cite{liu04}.

Co-occurrence or transactional data also appears in the context of
social networks, \emph{e.g.}, by considering which academic papers
are co-cited by another paper, which web pages are linked to or
from another web page, or which people were diagnosed with a
common disease on the same day.  Such measurements are readily
available, but do not necessarily reflect the temporal or other
natural order of occurrence.  Researchers in this area have
considered the problems of reconstructing networks from
co-occurrence data and of using the inferred network to predict
potential future co-occurrences \cite{kubica03}.

\emph{Functional magnetic resonance imaging} (fMRI) provides a
mechanism for measuring activity in the brain with high spatial
resolution.  By observing which regions of the brain co-activate
while a patient is performing different tasks we can obtain
multiple co-occurrence observations.  Although fMRI offers high
spatial resolution it has limited temporal resolution, making it
impractical to obtain complete order information.
Magnetoencephalography and electroencephalography measure activity
in the brain with higher temporal resolution but only provide
coarse spatial resolution, and thus may not allow one to determine
precisely which functional regions are active during a given task.
Existing techniques for obtaining functional co-activation
networks either involve brute-force measurement or use crude
correlation methods (see \cite{sporns02} and references therein).

In this article we focus on observations arising from transmissions in
a network.  Specifically, each co-occurrence observation corresponds
to a path\footnote{Throughout this paper a ``path'' refers to a
  sequence of vertices $(x_1, x_2, \dots, x_N)$ such that there is an
  edge between each adjacent pair of vertices, $x_{i-1}$ and $x_i$,
  and no node appears more than once in the sequence.}  through the
network. We observe the vertices comprising each path but not the
order in which they appear along the path.  In certain applications
the endpoints (source and destination) of the path may also be
observed.

Our goal is to identify which pairs of vertices are directly
connected via an edge, thereby learning the structure of the
network.  A \emph{feasible graph} is one which agrees with the
observations; \emph{i.e.}, a graph which contains a directed path
through the vertices in each co-occurrence observation.  Given a
collection of co-occurrence observations a feasible graph is
easily constructed by assigning an order -- any order, in fact --
to the vertices in each observation, and then inserting directed
edges between vertices which are adjacent in the assigned order.
In light of the many possible orders for each co-occurrence
observation, the number of feasible topologies grows exponentially
in the number and size of observations.  Without additional
assumptions, side information, or prior knowledge, there is no
reason to prefer one feasible topology over the others.

Previous work on related problems has involved heuristics using
frequencies of co-occurrence either to assign an order to each
path \cite{rabbat05} or to approximate the probability of
transitioning from one vertex to another \cite{kubica03}. These
approaches make stringent assumptions and sacrifice flexibility in
order to achieve computational tractability and systematically
identify a unique solution.  The \emph{frequency method}
introduced in \cite{rabbat05} is based on a model where paths from
a particular source or to a particular destination form a tree.
This model coincides with the shortest-path routing policy.  When
the network provides multiple paths between the same pair of
endpoints (\emph{e.g.}, for load-balancing) the algorithm may
fail.  The \emph{cGraph} algorithm of Kubica \emph{et
al.}~\cite{kubica03} inserts weighted edges between every pair of
vertices which co-occur in some observation.  This approach
produces solutions which are typically much denser than desired.
Because both of these methods are based on heuristics, the results
they produce are not easily interpreted.  Also, these heuristics
do not readily lend themselves to incorporating side information.
A different approach, introduced by Justice and Hero in
\cite{justice05}, involves averaging over an ensemble of feasible
topologies sampled uniformly from the feasible set.  In general
there is an enormous number of feasible topologies (exponential in
the problem dimensions) exhibiting a wide variety of
characteristics, and it is not clear that an average of feasible
topologies will be optimal in any sense. These observations have
collectively motivated our development of a more general approach
to network reconstruction which we simply term \emph{network
inference from co-occurrences}, or NICO for short.

Our approach is based on a generative model where paths are
realizations of a random walk on the underlying graph. A
co-occurrence observation is obtained by randomly shuffling each
path to account for our lack of observed order information.  Based
on this model, network inference reduces to estimating the
parameters governing the random walk.  Then, these parameter
estimates determine the most likely order for each co-occurrence.

The following interpretation motivates our shuffled random walk model.
Imagine sitting at a particular vertex in the network and observing a
series of transmissions pass by.  This vertex is only connected to a
handful of other vertices in the network, so regardless of its final
destination, a transmission arriving at this vertex must pass through
one of the neighboring vertices next.  Over a period of time, we could
record how many arriving transmissions are passed to each neighbor,
and then calculate the empirical probabilities of transmissions to
each of the neighbors.  Obtaining such probabilities at each vertex
would provide a tremendous amount of information about the network,
but unfortunately co-occurrence observations do not directly reveal
them and we therefore face a challenging inverse problem.  This paper
develops a formal framework for estimating local transition
probabilities from a collection of co-occurrence observations, without
making any additional assumptions about routing behavior or properties
of the underlying network structure.  Experimental results on
simulated topologies indicate that good performance is obtained for a
variety of operating conditions.

It is also worth mentioning that the approach discussed in this
paper differs considerably from that of learning the structure of
a directed graphical model or Bayesian network, a graph where
nodes correspond to random variables and edges indicate
conditional independence relationships
\cite{heckerman95,koller03}.  A typical aim of graphical modelling
is to find a graph corresponding to a factorization of a
high-dimensional distribution which predicts the observations
well.  In turn, these probabilistic models do not directly reflect
physical structures, and applying such an approach in the context
of this problem would ignore physical constraints inherent to the
observations: that co-occurring vertices must lie along a path in
the network.

The rest of the paper is organized as follows.  In
Section~\ref{sec:model} we introduce notation and formulate the
problem setup.  Section~\ref{sec:em} reviews the standard approach
to estimating the parameters of a random walk when fully observed
(ordered) samples are available and presents an EM algorithm for
estimating random walk parameters from shuffled observations.  A
Monte Carlo variant of the EM algorithm is described in
Section~\ref{sec:approxe} for situations where long transmission
paths make the exact E-step computation prohibitive.
Section~\ref{sec:convergence} analyzes convergence of the Monte
Carlo EM algorithm.  Simulation results are presented in
Section~\ref{sec:sims} and the paper is concluded in
Section~\ref{sec:conc}, where ongoing work is also briefly
described.

\section{Problem Formulation}
\label{sec:model}

We model the network as a simple directed graph $G=(S,E)$, where
$S = \{1,2,\dots,|S|\}$ is the set of vertices/nodes and
$E\subseteq S\times S$ is the set of edges. The number of
nodes, $|S|$, is considered known, so network inference
amounts to determining the adjacency structure of the graph;
that is, identifying whether or not
$(i,j)\in E$, for every pair of vertices $(i,j)\in S\times S$.

A co-occurrence observation, $\by \subset S$, is a subset of
vertices in the graph which simultaneously ``occur'' when a
particular stimulus is presented to the network.  For example,
when a transmission is made over a communication network, a subset
of routers and switches carry the transmission from the source to
the destination.  This activated subset corresponds to a
co-occurrence observation, with the stimulus being a transmission
between that particular source-destination pair.  By repeating
this procedure $T$ times with different stimuli we obtain
observations, $\mathcal{Y} = \{\by^{(1)},
\by^{(2)},\dots,\by^{(T)}\}$, where $\by^{(m)} = (y_1^{(m)}, y_2^{(m)},
\dots, y_{N_m}^{(m)})$ is a length-$N_m$ co-occurrence, indexed in
an arbitrary order.

A directed graph $G = (S,E)$ is said to be feasible with respect
to observations $\mathcal{Y}$ if for each co-occurrence $\by^{(m)}
\in \mathcal{Y}$ there exists an ordered path $\bz^{(m)} =
(z_1^{(m)}, z_2^{(m)}, \dots, z_{N_m}^{(m)})$ and a permutation
$\btau^{(m)} = (\tau_1^{(m)}, \dots, \tau_{N_m}^{(m)})$ such that
$z_t^{(m)} = y_{\tau_t^{(m)}}^{(m)}$ for each $t$, and there is an
edge from $z_{t-1}$ to $z_t$ in the graph for $t=2,\dots,N_m$,
that is, $(z_{t-1}, z_t) \in E$.

Notice that if we observed ordered paths then network inference
would be trivial.  We would begin with an empty graph $G_0 = (S,
E)$ with $E = \{\}$.  Then, for each ordered observation
$\bz^{(m)}$ we would update the set of edges via $E \leftarrow E
\cup (z_{t-1}^{(m)}, z_t^{(m)})$ for $t = 2,\dots, N_m$.
Similarly, if we observed the correct permutation $\btau^{(m)}$
along with each co-occurrence $\by^{(m)}$, we could invert the
permutation to recover ordered observations and apply the same
procedure.

In practice we do not make ordered observations nor do we have
access to the correct permutations.  However, we can obtain a
feasible reconstruction by associating \emph{any} permutation (of
the appropriate length) with each co-occurrence, and then
following the procedure described above. There are $N_m!$ ways to
permute the elements of $\by^{(m)}$, so there may be as many as
$\prod_{m=1}^{T} N_m!$ feasible reconstructions.  Clearly, for
large $N_m$ and $T$ this is a huge set to search over.  Moreover,
without making additional assumptions, or adopting some additional
criteria, there is no reason to prefer one feasible reconstruction
over another.

Physical principles governing the development of many natural and
man-made networks suggest that not all feasible networks are
equally plausible.  Intuitively, if two or more vertices appear
collectively in multiple co-occurrences, we expect that their
order is probably the same in the corresponding paths.  Likewise,
we expect that most vertices will only be directly connected to a
small fraction of the other vertices.  Based on this intuition we
propose the following probabilistic model.  First, we model the
unobserved, ordered paths, $\bz^{(m)}$, as independent samples of
a first-order Markov chain. The Markov chain is parameterized by
an initial state distribution $\bpi \in [0,1]^{|S|}$ where $\pi_i
= P[z_1 = i]$, and a probability transition matrix, $\Amat \in
[0,1]^{|S| \times |S|}$, where $A_{i,j} = P[z_t = j | z_{t-1} =
i]$. Of course, these parameters must satisfy the normalization
constraints
\begin{eqnarray}
  \sum_{i=1}^{|S|} \pi_i = 1 & \mbox{ and } & \sum_{j=1}^{|S|}
  A_{i,j} = 1,\hspace{0.2in} \mbox{ for each } i=1,\dots,|S|.
  \label{stochasticmatrix}
\end{eqnarray}
In addition, we assume that the support of the transition matrix
is determined by the adjacency structure of the underlying
network; \emph{i.e.}, $A_{i,j} > 0$ if and only if $(i,j) \in E$.

A co-occurrence observation, $\by$, is generated by shuffling the
elements of an ordered Markov chain sample, $\bz =
(z_1,\dots,z_N)$, via a permutation $\btau$ drawn uniformly from
$\Psi_N$, the collection of all permutations of $N$ elements.
Thus, for each $t=1,\dots,N$, $z_t = y_{\tau_t}$.  We assume that
$\btau$ is independent of the Markov chain sample, $\bz$.  Based
on this model, we can write the likelihood of a co-occurrence
observation $\by$ conditioned on the permutation $\btau$ as
\begin{eqnarray}
  P[\by | \btau, \Amat, \bpi] &=& \pi_{y_{\tau_1}} \prod_{t=2}^N
  A_{y_{\tau_{t-1}}, y_{\tau_{t}}}. \label{eq:condDist}
\end{eqnarray}
Since $P[\btau] = 1/(N!)$, for any $\btau\in \Psi_{N}$,
marginalization over all permutations leads to
\begin{equation}
  P[\by | \Amat, \bpi] = \frac{1}{N!} \sum_{\btau \in \Psi_N} P[\by | \btau, \Amat,
  \bpi].
\end{equation}
Finally, assuming  that co-occurrence observations are independent, and taking the logarithm,
gives
\begin{eqnarray}
  \log P[\mathcal{Y} | \Amat, \bpi] &=& \sum_{m=1}^T
  \left[\log\left(\sum_{\btau \in \Psi_{N_m}} P[\by^{(m)} |
  \btau^{(m)}, \Amat, \bpi]\right) - \log(N_m!)\right].
  \label{eq:marginal}
\end{eqnarray}
Under this model, network inference consists in computing the maximum
likelihood (ML) estimates,
\begin{equation}
  (\widehat{\Amat}_{\mbox{\scriptsize ML}}, \widehat{\bpi}_{\mbox{\scriptsize ML}}) =
  \arg \max_{\Amat, \bpi} \; \log P[\mathcal{Y} | \Amat, \bpi].
  \label{eq:mlopt}
\end{equation}

With the ML estimates in hand, we may determine the most likely permutation for each
co-occurrence observation according to $(\Amat_{\mbox{\scriptsize
ML}}, \bpi_{\mbox{\scriptsize ML}})$, and obtain a feasible
reconstruction using our procedure for ordered observations
described above.


For non-trivial observations, $\log P[\mathcal{Y} | \Amat, \bpi]$
is a complicated, non-concave function of $(\Amat, \bpi)$, so solving
(\ref{eq:mlopt}) is not a simple task.  In the next section, we derive
a EM algorithm for solving this optimization problem, by treating the
set of permutations, ${\cal T} = \{\btau^{(1)},...,\btau^{(T)}\}$,
shuffling the paths, as missing data.

\section{An EM Algorithm for Estimating Markov Chain Parameters from
Shuffled Observations} \label{sec:em}
\subsection{Fully Observed Markov Chains:
Notation and Estimation} \label{sec:mcdirect} Let $\mathcal{Z} =
\{\bz^{(1)},...,\bz^{(T)}\}$ be a set of sample paths, $\bz^{(m)}
= (z_1^{(m)},...,z_{N_m}^{(m)})$, independently generated by a
Markov chain with transition matrix $\Amat$ and initial state
distribution $\bpi$ (see (\ref{stochasticmatrix})). For later use,
it is convenient to introduce the equivalent binary representation
$\bw^{(m)} = (\bw_1^{(m)},...,\bw_{N_m}^{(m)})$, for each sample
$\bz^{(m)}$, defined as follows: $\bw_t^{(m)} =
(w_{t,1}^{(m)},...,w_{t,|S|}^{(m)}) \in \{0,1\}^{|S|}$, with $(w_{t,i}^{(m)} =
1) \Leftrightarrow (z_t^{(m)} = i)$; of course, one and only one
entry of each vector $\bw_t^{(m)}$ equals $1$. Finally, let
$\mathcal{W} = \{\bw^{(1)},...,\bw^{(T)}\}$, which contains the
exact same information as $\mathcal{Z}$. With this notation, we
can write
\begin{eqnarray}
\log P[{\cal W}|{\bf A},\bpi] & = &\sum_{m=1}^T \sum_{i=1}^{|S|}
w_{1,i}^{(m)} \log\pi_i + \sum_{m=1}^T \sum_{t=2}^{N_m}
\sum_{i=1}^{|S|} \sum_{j=1}^{|S|} w_{t-1,i}^{(m)}\, w_{t,j}^{(m)} \log A_{i,j}.\nonumber\\
& = & \sum_{i=1}^{|S|} \log\pi_i \sum_{m=1}^T  w_{1,i}^{(m)}
 + \sum_{i=1}^{|S|} \sum_{j=1}^{|S|} \log A_{i,j}
 \sum_{m=1}^T \sum_{t=2}^{N_m}  \, w_{t-1,i}^{(m)}\; w_{t,j}^{(m)} .\nonumber
\end{eqnarray}

Maximum likelihood estimates of $\bpi$ and ${\bf A}$ can be
obtained from $\mathcal{W}$ my maximizing $\log P[\mathcal{W}|{\bf
A},\bpi]$ under the constraints in (\ref{stochasticmatrix}); the
solution is well known,
\begin{equation}
  \widehat{A}_{i,j} = \frac{\displaystyle\sum_{m=1}^T \sum_{t=2}^{N_m}\;
  w_{t-1,i}^{(m)}\; w_{t,j}^{(m)}}{\displaystyle\sum_{j=1}^{|S|}\sum_{m=1}^T \sum_{t=2}^{N_m}
  \; w_{t-1,i}^{(m)}\; w_{t,j}^{(m)}},\hspace{1cm}\mbox{and}\hspace{1cm}
\widehat{\pi}_i = \frac{1}{T}\sum_{m=1}^T
w_{1,i}^{(m)}.\label{eq:mlestimates}
\end{equation}

\subsection{Shufflings, Permutations, and the EM Algorithm}
\label{sec:emalg}

To address the case where we have a set of co-occurrences
$\mathcal{Y} = \{\by^{(1)},\dots,\by^{(T)}\}$, not ordered
samples, we defined the equivalent binary representation
$\mathcal{X} = \{\bx^{(1)}, \dots, \bx^{(T)}\}$ for $\mathcal{Y}$
in a similar way as above: $\bx^{(m)} = (\bx_1^{(m)},...,
\bx_{N_m}^{(m)})$, where $\bx_t^{(m)} =
(x_{t,1}^{(m)},...,x_{t,|S|}^{(m)}) \in \{0,1\}^{|S|}$, with
$(x_{t,i}^{(m)} = 1) \Leftrightarrow (y_t^{(m)} = i)$.

Rather than using $\btau^{(m)} = (\tau_1^{(m)},\dots,\tau_N^{(m)})$
to denote the $m$th permutation/shuffling, we introduce a more convenient (binary)
representation; each shuffling is represented by a {\it
permutation matrix}\footnote{A matrix with one and only one ``1"
in each row and each column.}, which we will term {\it shuffling
matrix}. Let the shuffling matrix for sequence $m$ be denoted as
${\bf r}^{(m)}$ so that $(r_{t,t'}^{(m)}=1) \Leftrightarrow
(\tau_t = t') \Leftrightarrow (\bx^{(m)}_{t'} = \bw^{(m)}_t)
\Leftrightarrow (y^{(m)}_{t'} = z^{(m)}_t)$. Given both
$\br^{(m)}$ and $\bx^{(m)}$, we recover the  unshuffled sequence
${\bf w}^{(m)}$ by applying (using $0^0 = 1$)
\begin{equation}
w_{t,i}^{(m)} = \prod_{t'=1}^{N_m}
\left(x_{t',i}^{(m)}\right)^{r^{(m)}_{t,t'}}. \label{eq:unshuffle}
\end{equation}

Let ${\cal R} = \{{\bf r}^{(1)},...,{\bf r}^{(T)}\}$ be the
collection of shuffling matrices that allow recovering the
underlying ordered paths $\mathcal{W} =
\{\bw^{(1)},...,\bw^{(T)}\}$ from the corresponding shuffled
co-occurrences ${\cal X} = \{{\bf x}^{(1)},...,{\bf x}^{(T)}\}$.
We can write the complete log-likelihood $\log P[{\cal X},{\cal
R}|{\bf A},{\bpi}]$ as follows:
\begin{eqnarray}
\log P[{\cal X},{\cal R}|{\bf A},{\bpi}] & = & \log P[{\cal X}|{\cal R},{\bf A},{\bpi}] + \log p[{\cal R}]\\
& = & \sum_{m=1}^T \log P[{\bf x}^{(m)}|{\bf r}^{(m)},{\bf A},{\bpi}] + \log p[{\cal R}]\nonumber\\
& = & \sum_{m=1}^T \;\sum_{t=2}^{N_m}
\;\sum_{t'=1}^{N_m}\;\sum_{t''=1}^{N_m} \;\sum_{i,j=1}^{|S|}\;
r^{(m)}_{t,t'}\; r^{(m)}_{t-1,t''}\;
x_{t'',i}^{(m)}\; x_{t',j}^{(m)}\; \log A_{i,j} \nonumber\\
& & + \sum_{m=1}^T \;\sum_{t'=1}^{N_m}\; \sum_{i=1}^{|S|}\;
r^{(m)}_{1,t'}\; x_{t',i}^{(m)}\; \log \pi_i + \log p[{\cal
R}],\label{eq:complete}
\end{eqnarray}
where $p[{\cal R}]$ is the probability of the set of shufflings
${\cal R}$, which we assume constant.

To estimate ${\bf A}$ and $\bpi$ from $\mathcal{X}$, we treat
$\mathcal{R}$ as missing data, opening the door to the use of the
EM algorithm. Notice that if we had the complete data
$(\mathcal{X},\mathcal{R})$, we could recover $\mathcal{W}$ via
(\ref{eq:unshuffle}) and obtain the closed-form estimates
(\ref{eq:mlestimates}). The EM algorithm proceeds by computing the
expected value of $\log P[{\cal X},{\cal R}|{\bf A},{\bpi}]$
(w.r.t. $\mathcal{R}$), conditioned on the observations and on the
current model estimate $(\Amat^k,\bpi^k)$ (the E-step),
\begin{equation}
Q\left({\bf A},\bpi; \Amat^k, \bpi^k\right) = E\left[\log P[{\cal
X},{\cal R}|{\bf A},{\bpi}]\left| {\cal
X},\Amat^k,\bpi^k\right.\right].\label{eq:Estep}
\end{equation}
The model parameter estimates are then updated as follows (the
M-step):
\begin{equation}
\left(\Amat^{k+1},\bpi^{k+1}\right) = \arg\max_{{\bf A},\bpi}\;
Q\left({\bf A},\bpi; \Amat^k, \bpi^k\right). \label{eq:Mstep}
\end{equation}
These two steps are repeated cyclically until a convergence
criterion is met.

\subsection{The E-step}
\label{sec:Estep}
\subsubsection{Sufficient statistics}
Rearranging  \eqref{eq:complete}, and dropping $\log
P[\mathcal{R}]$ (assumed constant), we can write
\begin{eqnarray}
  \log P[\mathcal{X}, \mathcal{R} | \Amat, \bpi] &\propto&
  \sum_{m=1}^T \sum_{i,j=1}^{|S|} \;\; \sum_{t',t''=1}^{N_m}\;\;
  \sum_{t=2}^{N_m} r_{t,t'}^{(m)} r_{t-1,t''}^{(m)}
  x_{t'',i}^{(m)} x_{t',j}^{(m)} \log A_{i,j}\nonumber\\
  & & + \; \sum_{m=1}^T \sum_{i=1}^{|S|} \sum_{t'=1}^{N_m}
  r_{1,t'}^{(m)} x_{t',i}^{(m)} \log \pi_i.\label{eq:complete3}
\end{eqnarray}
revealing that $\log P[\mathcal{X}, \mathcal{R} | \Amat, \bpi]$ is
linear with respect to simple functions of $\mathcal{R}$:
\begin{itemize}
\item the first row of each ${\bf r}^{(m)}$: $r^{(m)}_{1,t'}$, for
$m=1,\dots,T$ and $t'=1,\dots,N_m$; \item sums of transition
indicators: $\alpha^{(m)}_{t',t''} \equiv \sum_{t=2}^{N_m}
r^{(m)}_{t,t'}\; r^{(m)}_{t-1,t''}$, for $m=1,\dots,T$, and
$t',t''=1,\dots,N_m$.
\end{itemize}
Since expectations commute with linear functions, the E-step
reduces to computing the conditional expectations of
$r^{(m)}_{t,t'}$ and $\alpha^{(m)}_{t',t''}$ (denoted
$\bar{r}^{(m)}_{1,t'}$ and $\bar{\alpha}^{(m)}_{t',t''}$,
respectively) and plugging them into the complete log-likelihood.
Noticing that the $r^{(m)}_{t,t'}$ and $\alpha^{(m)}_{t',t''}$ are
binary (in $\{0,\, 1\}$), yields
\begin{eqnarray}
\bar{r}^{(m)}_{1,t'}  & =  & E\left[r^{(m)}_{1,t'}\left| {\cal
X},\Amat^k,\bpi^k\right.\right] \; = \;
P\left[r^{(m)}_{1,t'}=1\left|
{\cal X},\Amat^k,\bpi^k\right.\right]\label{Estep1}\\
\bar{\alpha}^{(m)}_{t',t''} &  = &
E\left[\alpha^{(m)}_{t',t''}\left| {\cal
X},\Amat^k,\bpi^k\right.\right] \; =
P\left[\alpha^{(m)}_{t',t''}=1\left| {\cal
X},\Amat^k,\bpi^k\right.\right].\label{Estep2}
\end{eqnarray}
Finally, $Q\left({\bf A},\bpi;\Amat^k,\bpi^k\right)$ is obtained
simply by plugging $\bar{r}^{(m)}_{1,t'}$ and $\bar{\alpha}^{(m)}_{t',t''}$
in the places of $r^{(m)}_{1,t'}$ and $\sum_{t=2}^{N_m}
r^{(m)}_{t,t'}\; r^{(m)}_{t-1,t''}$, respectively,  in
(\ref{eq:complete3}).

\subsubsection{Computing $\bar{r}^{(m)}_{1,t'}$}
Since the permutations are (a priori) equiprobable, we have
$P[{\bf r}] = 1/(N_m!)$ (for ${\bf r}\in \Psi_{N_m}$),
$P\big[r_{1,t'}^{(m)}=1] = ((N_m - 1)!/N_m !) = 1/N_m$, and $P[{\bf
r}|r_{1,t'}^{(m)}=1] = 1/((N_m - 1)!)$. Using these facts,
together with the mutual independence among the several sequences,
and Bayes law, yields
\begin{eqnarray}
\bar{r}^{(m)}_{1,t'}  & = & P\left[r^{(m)}_{1,t'}=1\left|
{\bf x}^{(m)},\Amat^k,\bpi^k\right.\right]\nonumber\\
& = &  \frac{P\big[\bx^{(m)} \big| r_{1,t'}^{(m)}=1, \Amat^k,
\bpi^k\big]\;
P\big[r_{1,t'}^{(m)}=1]}{P[\bx^{(m)} \big| \Amat^k, \bpi^k\big]}\nonumber\\
& = & \frac{\displaystyle\sum_{\br \in \Psi_{N_m}\! :\;  r_{1,t'}
= 1} P\big[ \bx^{(m)} \big| \br, \Amat^k,
\bpi^k\big]}{\displaystyle\sum_{\br \in \Psi_{N_m}}
P\big[\bx^{(m)} \big| \br, \Amat^k,
\bpi^k\big]},\label{eqn:gamma1}
\end{eqnarray}
where each term $P\big[\bx^{(m)} \big| \br, \Amat^k, \bpi^k\big]$
is easily computed after using $\br$ to unshuffle $\bx^{(m)}$:
\[
  P\big[\bx^{(m)} \big| \br, \Amat^k, \bpi^k\big] =
  P\big[{\bf y}^{(m)} \big| \btau, \Amat^k, \bpi^k\big]
  = \pi^k_{y^{(m)}_{\tau_1}} \prod_{t=2}^{N_m} A^k_{y^{(m)}_{\tau_{t-1}},
  y^{(m)}_{\tau_t}}.\nonumber
\]
Denoting the numerator of (\ref{eqn:gamma1}) as
$\gamma_{t'}^{(m)}$ we have a  more compact expression
\[
\bar r_{1,t'}^{(m)} = \frac{\displaystyle\gamma_{t'}^{(m)}}{
\displaystyle\sum_{t'=1}^{N_m}
\gamma_{t'}^{(m)}}.
\]

\subsubsection{Computing $\bar \alpha_{t',t''}^{(m)}$}
The computation of $\bar \alpha_{t',t''}^{(m)}$ follows a similar
path as that of $\bar r_{1,t'}^{(m)}$; since all permutations are equiprobable,
$P[r_{t,t'}^{(m)} r_{t-1,t''}^{(m)}=1] = (N_m-2)!/(N_m!)$ and
$P[\br | r_{t,t'}^{(m)} r_{t-1,t''}^{(m)}=1] = 1/((N_m-2)!)$, thus
\begin{eqnarray}
\bar \alpha_{t',t''}^{(m)} &=& \sum_{t=2}^{N_m}
P\big[r_{t,t'}^{(m)} r_{t-1,t''}^{(m)}=1
\big| \bx^{(m)}, \Amat^k, \bpi^k\big] \nonumber\\
&=& \sum_{t=2}^{N_m} \frac{P\big[\bx^{(m)} \big| r_{t,t'}^{(m)}
r_{t-1,t''}^{(m)} =1, \Amat^k, \bpi^k\big]\; P\big[r_{t,t'}^{(m)}
r_{t-1,t''}^{(m)} = 1\big]}{P\big[\bx^{(m)} \big| \Amat^k, \bpi^k\big]}\nonumber\\
&=& \sum_{t=2}^{N_m} \frac{\displaystyle\left(\frac{1}{(N_m -2)!} \sum_{\br \in
\Psi_{N_m} : r_{t,t'} r_{t-1,t''} = 1} P\big[\bx^{(m)} \big| \br,
\Amat^k, \bpi^k\big]\right) \left(\frac{(N_m
-2)!}{N_m!}\right)}{\displaystyle\frac{1}{N_m!} \sum_{\br \in \Psi_{N_m}}
P\big[ \bx^{(m)} \big| \br, \Amat^k,
\bpi^k\big]}\nonumber\\
&=& \frac{\displaystyle\sum_{\br \in \Psi_{N_m}} P[\bx^{(m)} | \br, \Amat^k,
\bpi^k] \sum_{t=2}^{N_m} r_{t,t'} r_{t-1,t''}}{\displaystyle\sum_{\br \in
\Psi_{N_m}} P[\bx^{(m)} | \br, \Amat^k,
\bpi^k]}.\label{eqn:gamma3}
\end{eqnarray}
Denoting the numerator of (\ref{eqn:gamma3}) as
$\gamma_{t',t''}^{(m)}$, we finally have
\begin{equation}
  \bar \alpha_{t',t''}^{(m)} =
  \frac{\displaystyle\gamma_{t',t''}^{(m)}}{\displaystyle\sum_{t'=1}^{N_m}
  \gamma_{t',t''}^{(m)}}.
\end{equation}

For the $m$th observation, the statistics $\{\bar
r_{1,t'}^{(m)}\}$ and $\{\bar \alpha_{t,t',t''}^{(m)}\}$ have an
$O(N_m^2)$ memory cost ($ N_m^2 - N_m$ transition statistics and
$N_m$ initial state statistics).  These quantities can be computed
via the summary statistics $\{\gamma_{t'}^{(m)}\}$ and
$\{\gamma_{t',t''}^{(m)}\}$, using the same memory needed to store
$\{\bar r_{1,t'}^{(m)}\}$ and $\{\bar \alpha_{t,t',t''}\}$, in
$O\big(N_m !\big)$ operations (the number of all permutations for
a length $N_m$ observation). For large   $N_m$, this is a heavy
load; Section~\ref{sec:approxe}  describes a sampling approach for
computing approximations to $\bar r_{1,t'}$ and $\bar
\alpha_{t',t''}$.

\subsection{The M-step}
Recall that the function $Q\left({\bf
A},\bpi;\Amat^k,\bpi^k\right)$ is obtained by plugging $\bar
r_{1,t'}$ and $\bar \alpha_{t',t''}$ in the places of
$r^{(m)}_{1,t'}$ and $\sum_{t=2}^{N_m} r^{(m)}_{t,t'}\;
r^{(m)}_{t-1,t''}$, respectively, in (\ref{eq:complete3}).
Maximization w.r.t. ${\bf A}$ and $\bpi$, under the constraints in
\eqref{stochasticmatrix}, leads to following simple update
equations:
\begin{equation}
  A^{k+1}_{i,j} =
  \frac{\displaystyle \sum_{m=1}^T \sum_{t',t''=1}^{N_m}
  \bar \alpha_{t',t''}^{(m)} x_{t'',i}^{(m)}
  x_{t',j}^{(m)}}{\displaystyle\sum_{j=1}^{|S|}\sum_{m=1}^T \sum_{t',t''=1}^{N_m}
  \bar \alpha_{t',t''}^{(m)} x_{t'',i}^{(m)}
  x_{t',j}^{(m)}}\;\;\; \mbox{and} \;\;\;
\pi_i^{k+1} = \frac{\displaystyle\sum_{m=1}^T
\;\sum_{t'=1}^{N_m}\; \bar{r}^{(m)}_{1,t'}\; x_{t',i}^{(m)}
}{\displaystyle\sum_{i=1}^{|S|}\sum_{m=1}^T \;\sum_{t'=1}^{N_m}\;
\bar{r}^{(m)}_{1,t'}\; x_{t',i}^{(m)}}. \label{eqn:Mstep_A}
\end{equation}

\subsection{Handling Known Endpoints}
In some applications, (one or both of) the endpoints of each path
are known and only the internal nodes are shuffled. This is the
case in communication networks ({\it i.e.}, internally-sensed
network tomography), since the sources and destinations  are
known, but not the connectivity within the network. In estimation
of biological networks (signal transduction pathways), a physical
stimulus (\emph{e.g.}, hypotonic shock) causes a sequence of
protein interactions, resulting in another observable physical
response (\emph{e.g.}, a change in cell wall structure); in this
case, the stimulus and response act as fixed endpoints, our goal
is to infer the order of the sequence of protein interactions.

Observe that knowledge of the endpoints of each path imposes the
constraints
\[
r_{1,1}^{(m)} = 1 \hspace{0.2in} \mbox{and} \hspace{0.2in}
r_{N_m,N_m}^{(m)} = 1.
\]
Under the first constraint, estimates of the initial state
probabilities are simply given by
\[
  \widehat \pi_i = \frac{1}{T}\sum_{m=1}^T x_{1,i}^{(m)}.
\]
Thus, EM  only needs to be used to estimate the transition matrix
entries. Let
\[
  \widetilde \Psi_N = \{r \in \Psi_N\ :\ r_{1,1}=1,\, r_{N,N}=1\},
\]
denote the set of permutations of $N$ elements with fixed
endpoints.  As in the general case, the E-step can be computed
using summary statistics (for $t',t''=1,\dots,N_m$)
\begin{eqnarray}
  \widetilde \gamma^{(m)} & = & \sum_{r \in \widetilde \Psi_{N_m}} P[\bx^{(m)} |
  \br, \widehat{\Amat}, \widehat{\bpi}]\nonumber\\
  \widetilde \gamma_{t',t''}^{(m)} & = & \sum_{r \in \widetilde \Psi_{N_m}}
  \! P[\bx^{(m)} | \br, \widehat{\Amat}, \widehat{\bpi}] \;\;
  \sum_{t=2}^{N_m} r_{t,t'} r_{t-1,t''},\nonumber
\end{eqnarray}
and setting $\bar \alpha_{t',t''}^{(m)} = \widetilde
\gamma_{t',t''}^{(m)} / \widetilde \gamma$.
 The M-step (update for $\Amat^{k+1}$) remains unchanged.

 \subsection{Incorporating Prior Information}
\label{sec:priorinfo} The EM algorithm can be easily modified to
incorporate conjugate  priors; these are Dirichlet priors for
$\bpi$ and for each row of $\Amat$,
\begin{equation}
  P[\bpi | {\bf u}] \propto \prod_{i=1}^{|S|} \pi_i^{u_i - 1}
  \hspace{1cm} \mbox{and} \hspace{1cm}
  P[\Amat | {\bf v}] \propto \prod_{i=1}^{|S|}\prod_{j=1}^{|S|}
  A_{i,j}^{v_{i,j}-1}, \label{eqn:transprior}
\end{equation}
where the parameters $u_i$ and $v_{i,j}$ should be non-negative in
order to have proper priors \cite{bernardo94}.  The larger $u_i$
is relative to the other $u_{i'}$, $i' \ne i$, the greater our
prior belief that state $i$ is an initial state rather than the
others. Similarly, the larger $v_{i,j}$ relative to other
$v_{i,j'}$ for $j' \ne j$, the more likely we expect, \emph{a
priori}, transitions from state $i$ to state $j$ relative to
transitions from $i$ to the other states.

Adding the logarithms of the priors in (\ref{eqn:transprior}) to
the complete log-likelihood (\ref{eq:complete}), we find that
incorporating priors into the EM algorithm only results in a
change to the M-step. Consider the prior distribution on the
initial state distribution; taking $u_i = c > 1$, for all $i$, has
a {\it smoothing} effect, encouraging all of the states to have
some mass in the initial state distribution. On the other hand,
with $0 < c < 1$  will have a {\it shrinkage} effect, encouraging
all of the mass to go to one (or a few) of the states. We can push
even more aggressively for a sparse solution by choosing negative
parameters for the Dirichlet distributions (which will become
improper), as done in \cite{figueiredo02} for Gaussian mixtures.
When negative Dirichlet parameters are allowed, the M-step updates
become
\begin{eqnarray}
   \pi_i^{t+1} &=& \frac{\displaystyle\left(u_i +
  \sum_{m=1}^T \sum_{t'=1}^{N_m} \bar r^{(m)}_{1,t'}\right)_+}{\displaystyle\sum_{i=1}^{|S|} \left(u_i +
  \sum_{m=1}^T \sum_{t'=1}^{N_m} \bar r^{(m)}_{1,t'}\right)_+}\\
  A_{i,j}^{k+1} &=&
  \frac{\displaystyle\left(v_{i,j} + \sum_{m=1}^T \sum_{t'=1}^{N_m}
  \sum_{t''=1}^{N_m} \bar \alpha_{t',t''}^{(m)} x_{t-1,i}^{(m)}
  x_{t,j}^{(m)}\right)_+}{\displaystyle\sum_{j=1}^{|S|}\left(v_{i,j} + \sum_{m=1}^T \sum_{t'=1}^{N_m}
  \sum_{t''=1}^{N_m} \bar \alpha_{t',t''}^{(m)} x_{t-1,i}^{(m)}
  x_{t,j}^{(m)}\right)_+}.
\end{eqnarray}
where $(a)_+ = \max\{0,a\}$ is the so-called {\it positive part}
operator.

\section{Monte Carlo E-Step by Importance Sampling}
\label{sec:approxe}

For long sequences, the combinatorial nature of (\ref{eqn:gamma1})
and (\ref{eqn:gamma3}) (involving sums over all permutations of
each sequence) may render exact computation impractical. In this
section, we consider Monte Carlo approximate versions of the
E-step, which avoid the combinatorial nature of its exact version.
The Monte Carlo EM (MCEM) algorithm, based on an MC version of the
E-step, was originally proposed in \cite{wei90}, and used ever since
by many authors (recent work can be found in \cite{jank05,booth01}
and references therein).

To lighten the notation in this section, we drop the superscripts
from $(\Amat^k,\, \bpi^k)$, using simply $(\Amat,\bpi)$ as the
current parameter estimates. Moreover, we focus on a particular
length-$N$ path $\by = \{\by_1,\dots,\by_N\} \subseteq S^N$ and
drop the superscript $(m)$; due to the independence of the paths,
there's no loss of generality. Recall that $\by$ is a (shuffled)
path, thus has no repeated elements.

The E-step (see (\ref{Estep1}) and (\ref{Estep2})) consists of
computing the conditional expectations $ \bar r_{1,t'} =
E\big[r_{1,t'} \big| \bx,
  \Amat, \bpi \big]$ and $\bar \alpha_{t',t''} = E\big[\alpha_{t',t''} \big| \bx, \Amat,
  \bpi\big] $.
A na{\"\i}ve  Monte Carlo approximation would be based on random
permutations, sampled from the uniform distribution over $\Psi_N$.
However, the reason to resort to approximation techniques in the
first place is that $\Psi_N$ is large, with only a small fraction
of these random permutations having non-negligible posterior
probability, $P[\br | \bx, \Amat, \bpi]$; a very large number of
uniform samples is thus needed to obtain a good approximation to
$\bar r_{1,t'}$ and $\bar \alpha_{t',t''}$.

Ideally, we would sample permutations directly from the posterior
 $P[\br | \bx, \Amat, \bpi]$; however, this would
require determining its value for all $N!$ permutations. Instead,
we employ \emph{importance sampling} (IS) (see, {\it e.g.},
\cite{RobertCasella,liu01},
 for an introduction to IS): we sample $L$ permutations, $\br^1,\dots,\br^L$,
from a distribution $R[\br| \bx, \Amat, \bpi]$, from  which it is easier to sample
than $P[\br | \bx, \Amat, \bpi]$, then apply a corrective
re-weighting to obtain approximations to $\bar r_{1,t'}$ and $\bar
\alpha_{t',t''}$ (note that we are now using superscripts on $\br$
to index sample numbers, not to identify paths). The IS estimates
are given by
\begin{eqnarray}
  \bar r_{1,t'} &\simeq& \frac{\displaystyle\sum_{i=1}^L z_i\;
  r_{1,t'}^i}{\displaystyle\sum_{i=1}^L z_i}, \label{eqn:is_r}\\
  \bar \alpha_{t,t',t''} &\simeq& \frac{\displaystyle\sum_{i=1}^L z_i\;
  \sum_{t=2}^{N_m} r_{t,t'}^i\; r_{t-1,t''}^i}{\displaystyle\sum_{i=1}^L z_i},
  \label{eqn:is_alpha}
\end{eqnarray}
where $z_i$,  the correction factor (or weight) for sample
$\br^i$, is given by
\begin{equation}
  z_i = \frac{P[\br^i | \bx, \Amat, \bpi]}{R[\br^i| \bx, \Amat, \bpi]},
\end{equation}
the ratio between the desired distribution and the sampling
distribution employed.

A relevant observation is that the target and sampling
distributions only need to be known up to normalizing factors.
Given $R'[\br] = Z_R\, R[\br| \bx, \Amat, \bpi]$ and $P'[\br | \bx, \Amat, \bpi] =
Z_P\, P[\br | \bx, \Amat, \bpi]$, for constants $Z_R$ and $Z_P$,
we can use
\begin{equation}
  z'_i = \frac{P'[\br^i | \bx, \Amat, \bpi]}{R'[\br^i| \bx, \Amat, \bpi]}
   = \frac{Z_P}{Z_R}\; z_i,
\end{equation}
instead of $z_i$ in \eqref{eqn:is_r} and \eqref{eqn:is_alpha};
these sums will remain unchanged since the factor $Z_P / Z_R$ will
appear both in the numerator and denominator, thus cancelling
out.

The remainder of this section contains the description of an IS
scheme, including the derivation of closed form expressions for
both the sampling distribution, $R$, and the sample weights,
$z_i$. We conclude the section by mentioning other sampling
variants and presenting experimental results.

\subsection{Sampling Scheme}
\label{subsec:causalIS}

Let  ${\bf f} = \{f_1,\dots,f_{|S|}\} \in \{0,1\}^{|S|}$ be a
sequence of binary flags.   Given some probability distribution
$\bp = \{p_1, p_2, \dots, p_{|S|}\}$ on the set of states, $S$,
denote by $\bp \cdot {\bf f}$ the restriction of $\bp$ to those
elements of $S$ that have corresponding flag $f_i$ set to 1, that
is,
\begin{equation}
  (\bp \cdot {\bf f})_i = \frac{\displaystyle p_i\; f_i}{\displaystyle
  \sum_{j=1}^{|S|} p_j \; f_j} , \hspace{0.2in} \mbox{for}\ i=1,2,\dots,|S|.
  \label{eq:masking}
\end{equation}

The proposed sampling scheme is defined as follows:
\begin{description}
\item[Step 1:] Let ${\bf f} = \{f_1,\dots,f_{|S|}\}$ be
initialized according to  $f_i = \1{i \in \by},$ where $\1{}$
denotes the indicator function.

Obtain one sample from $S$ according to the distribution $\bpi
\cdot {\bf f}$.  Let the obtained sample be denoted $s$; of
course, one and only one element of $\by$ is equal to $s$.

Locate the position $t$ of $s$ in $\by$; that is, find $t$ such
that $y_t = s$. Set $\tau_1 = t$.

Set $f_s = 0$ (preventing $y_t$ from being sampled again). Set
$i=2$.

\item[Step 2:] Let $\bp = \{A_{s,1}, \dots, A_{s,|S|}\}$ be the
$s$th row of the transition matrix.

Obtain a sample $s'$ from $S$, according to the distribution $\bp
\cdot {\bf f}$.

Find $t$ such that $y_t = s'$. Set $\tau_i = t$. Set $f_{s'} = 0$.

\item[Step 3:] If $i < N$, then set $s \leftarrow s'$, set  $i
\leftarrow i+1$, go back to Step 2; otherwise, stop.
\end{description}

\subsubsection{Sampling Distribution}
\label{sec:samplingdist1} Before deriving the form of the
distribution $R$, let us begin by writing the target distribution
$P[\btau | \by, \Amat, \bpi]$ explicitly.  Using Bayes law,
\begin{equation}
  P[\btau | \by, \Amat, \bpi] = \frac{P[\by |
  \btau, \Amat, \bpi] P[\btau]}{P[\by |
  \Amat, \bpi]}, \label{eqn:permPosterior}
\end{equation}
since $\btau$ does not depend \emph{a priori} on $\Amat$ or
$\bpi$. Based on our assumption that all permutations are
equiprobable we have $P[\btau] = \1{\btau \in \Psi_N} / N!$.
Noticing that the denominator in \eqref{eqn:permPosterior} is just
a normalizing constant independent of $\btau$, we have
\begin{equation}
  P[\btau | \by, \Amat, \bpi] \propto \1{\btau
  \in \Psi_N}\ P[\by | \btau, \Amat, \bpi] =
  \1{\btau \in \Psi_N} \left( \pi_{y_{\tau_1}}
  \prod_{t=2}^N A_{y_{\tau_{t-1}}, y_{\tau_t}} \right).
  \label{eqn:desiredDist}
\end{equation}

For the sake of notational economy, we will write simply
$R[\btau]$ to represent $R[\btau| \by, \Amat, \bpi]$. The sequential
nature of the sampling scheme suggests a factorization of the form
\begin{equation}
  R[\btau] = R[\tau_1] \; R[\tau_2 | \tau_1] \; R[\tau_3 | \tau_2,
  \tau_1] \; \cdots \; R[\tau_N | \tau_{N-1},\dots,\tau_1].
  \label{eqn:seqISfactorization}
\end{equation}

For Step~1 of the sampling scheme, it's clear that, for
$\tau_1=1,\dots,N$,
\begin{equation}
  R[\tau_1] \propto \pi_{y_{\tau_1}}.
  \label{eqn:seqIS1}
\end{equation}
For the $i$-th iteration, we have,
\begin{equation}
  R[\tau_i | \tau_{i-1},\dots,\tau_1] =
  A_{y_{\tau_{i-1}},y_{\tau_i}}\; \phi_i(\tau_{i-1},\dots,\tau_1)\;
  \1{\tau_i \notin \{\tau_{i-1},\dots,\tau_1\}},
  \label{eqn:seqIS3}
\end{equation}
with
\[
  \phi_i(\tau_{i-1},\dots,\tau_1) = \left(\sum_{t \notin
  \{\tau_{i-1},\dots,\tau_1\}} A_{y_{\tau_{i-1}}, y_t}\right)^{-1}.
\]

Inserting \eqref{eqn:seqIS1}  and \eqref{eqn:seqIS3} into
\eqref{eqn:seqISfactorization}, we finally have
\begin{equation}
  R[\btau] \propto \left[\pi_{y_{\tau_1}} \prod_{i=2}^N
  A_{y_{\tau_{i-1}},y_{\tau_i}} \right] \left[\prod_{i=2}^N
  \phi_i(\tau_{i-1},\dots,\tau_1)\right] \left[\prod_{i=2}^N
  \1{\tau_i \notin \{\tau_{i-1},\dots,\tau_1\}}\right];
  \label{eqn:seqISsamplingDist}
\end{equation}
observe that the third factor in the r.h.s. of
(\ref{eqn:seqISsamplingDist}) is simply the indicator that $\btau$
is a permutation, {\it i.e.}, is equal to $\1{\btau \in \Psi_N}$,
for any $\btau \in\{1,...,N\}^N$.

Dividing \eqref{eqn:desiredDist} by \eqref{eqn:seqISsamplingDist}
we obtain the correction factor $z$ for a permutation sample
$\btau$ generated using this sequential scheme as
\begin{equation}
  z = \left(\prod_{i=2}^N
  \phi_i(\tau_{i-1},\dots,\tau_1)\right)^{-1} = \prod_{i=2}^N
  \;\;\sum_{t \notin \{\tau_{i-1},\dots,\tau_1\}}
  A_{y_{\tau_{i-1}},y_t}.
\end{equation}
With this quantity in hand, we have all the ingredients needed to
produce IS estimates of $\bar r_{1,t'}$ and $\bar \alpha_{t,t',t''}$.
Notice that computing the terms $\phi_i$, thus computing $z$, is
easy since these factors are the normalization terms for the
distributions $\bp \cdot {\bf f}$, which are already computed
while performing each iteration of Step~2.  Thus, we just need to
store the product of these normalizing constants to finally obtain
the weight $z$.

\subsubsection{Known Endpoints}
In the case where the endpoints are known, we fix $\tau_1 = 1$,
$\tau_N = N$, and set $f_1 = 0$ and $f_N = 0$ in Step 1; the
remainder of the procedure is unchanged. Based on these
constraints, the importance sampling weight takes a slightly
different form:
\begin{equation}
  z = \pi_{y_1} \; A_{y_{\tau_{N-1}}, y_N}\;  \prod_{i=2}^{N-1} \sum_{t \notin
  \{\tau_{i-1},\dots,\tau_1\}} A_{y_{\tau_{i-1}},y_t}
  .
\end{equation}


\subsection{Hierarchical Sampling Schemes}
\label{subsec:twoStageIS} In addition to the sampling scheme that
we have just described, we have also developed other sampling
schemes that work in a hierarchical, rather than sequential,
fashion. For the sake of space, we refrain from describing these
other sampling schemes; detailed descriptions can be found in
\cite{RabbatFigueiredoNowak_TR}. In particular, we have developed
a two-stage hierarchical scheme and a fully hierarchical scheme.
In the two-stage method, the first stage samples from the
collection of all possible transitions occurring in a path; then
the second stage samples from  the distribution on all
arrangements of these transitions, to form a permutation. In the
fully hierarchical method, the first stage samples a suitable set
of transitions, say ${\cal G}_1$; then, the following stage
samples a suitable collection of pairs of elements of ${\cal
G}_1$, yielding a collection of quadruples, ${\cal G}_2$, and the
procedure is repeated until a permutation is obtained.

\subsection{Performance Assessment}
A standard error metric for comparing two distributions $P$ and
$\widehat{P}$ taking values on the finite set $\Psi_N$ is the
$\ell_1$ distance, defined as
\begin{equation}
  \|P - \widehat{P}\|_{1} = \sum_{\br \in \Psi_N} \left| P[\br] -
  \widehat{P}[\br]\right|.
\end{equation}
Given a set of permutations, $\{\br^1,\dots,\br^L\}$, drawn from
the sampling distribution $R$ along with the corresponding
weights, $z_1, \dots, z_L$, we can compute the empirical
distribution $\widehat{P}_R$ induced on $\Psi_N$ according to
\begin{equation}
  \widehat{P}_R[\br] = \left(\sum_{i=1}^L z_i\right)^{-1} \sum_{i=1}^L \; z_i\; \1{\br^i = \br}.
\end{equation}
Notice that the Monte Carlo sufficient statistics $\widehat
\alpha_{t',t''}^{(m)}$ and $\widehat r_{1,t'}^{(m)}$ are just sums
of terms $\widehat{P}_R[\br]$, {\it e.g.}, $\widehat
\alpha_{t',t''} = \sum_{\br \in \Psi_N} \widehat{P}_R[\br]
\sum_{t=2}^{N_m} r_{t-1,t''} r_{t,t'}$.  Thus,
\[
\left|\bar \alpha_{t',t''}^{(m)} - \widehat
\alpha_{t',t''}^{(m)}\right| \le \left\|P -
\widehat{P}_R\right\|_1 \hspace{1cm} \mbox{and} \hspace{1cm}
\left|\bar r_{1,t'}^{(m)} - \widehat r_{1,t'}^{(m)}\right| \le
\left\|P - \widehat{P}_R\right\|_1,
\]
showing that if the $\ell_1$ error between the true distribution
on permutations and the empirical importance sampling distribution
is small, then all of the estimated sufficient statistics will be
close to the corresponding exact value.

We have evaluated the performance of various sampling schemes via
simulation, considering three scenarios concerning the
distribution over all permutations: 1) roughly uniform, 2)
moderately peaked, and 3) highly concentrated around just a few
permutations.  The scenario 1) is typical of the first EM
iterations; scenario 2) is typical during intermediate EM
iterations; the third scenario is typical when EM is near
convergence. We consider a length-8 path with known endpoints,
thus with $6! = 720$ possible orderings. This length is long
enough to suggest how each sampling scheme will perform for longer
paths, while still allowing enumeration of all orderings.

\begin{figure*}
\centering %
\subfigure[]{\includegraphics[width=2.1in]{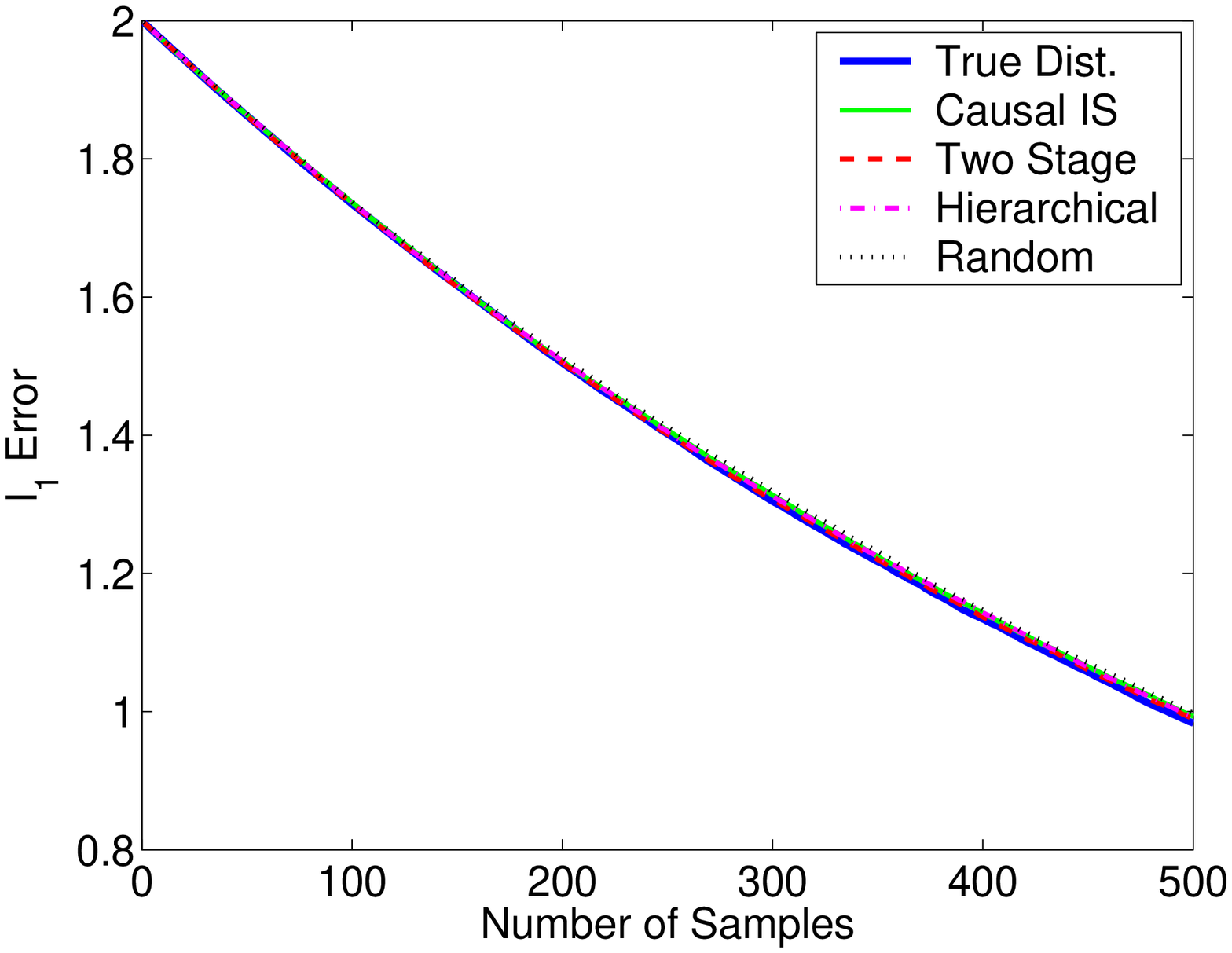}} %
\subfigure[]{\includegraphics[width=2.2in]{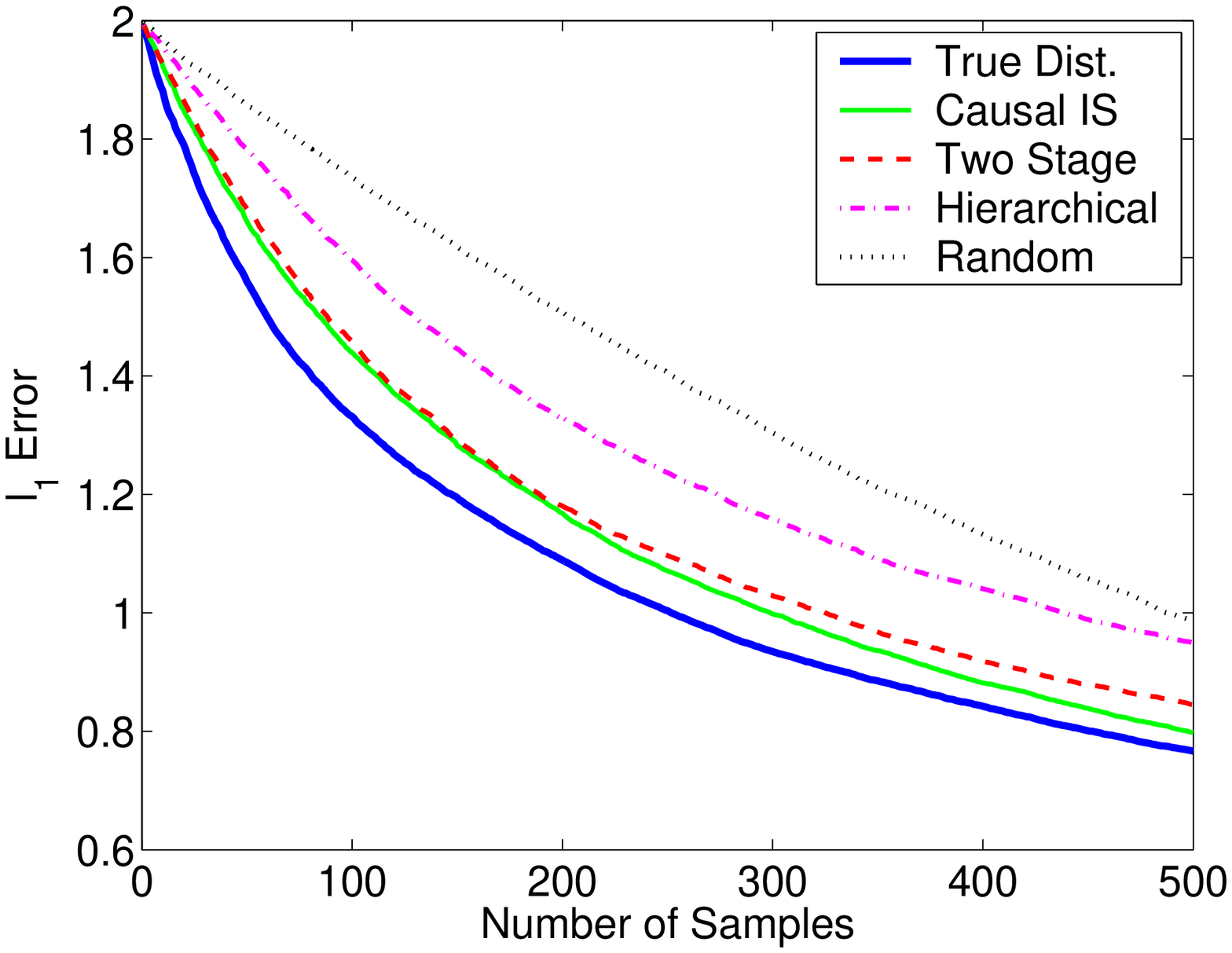}} %
\subfigure[]{\includegraphics[width=2.1in]{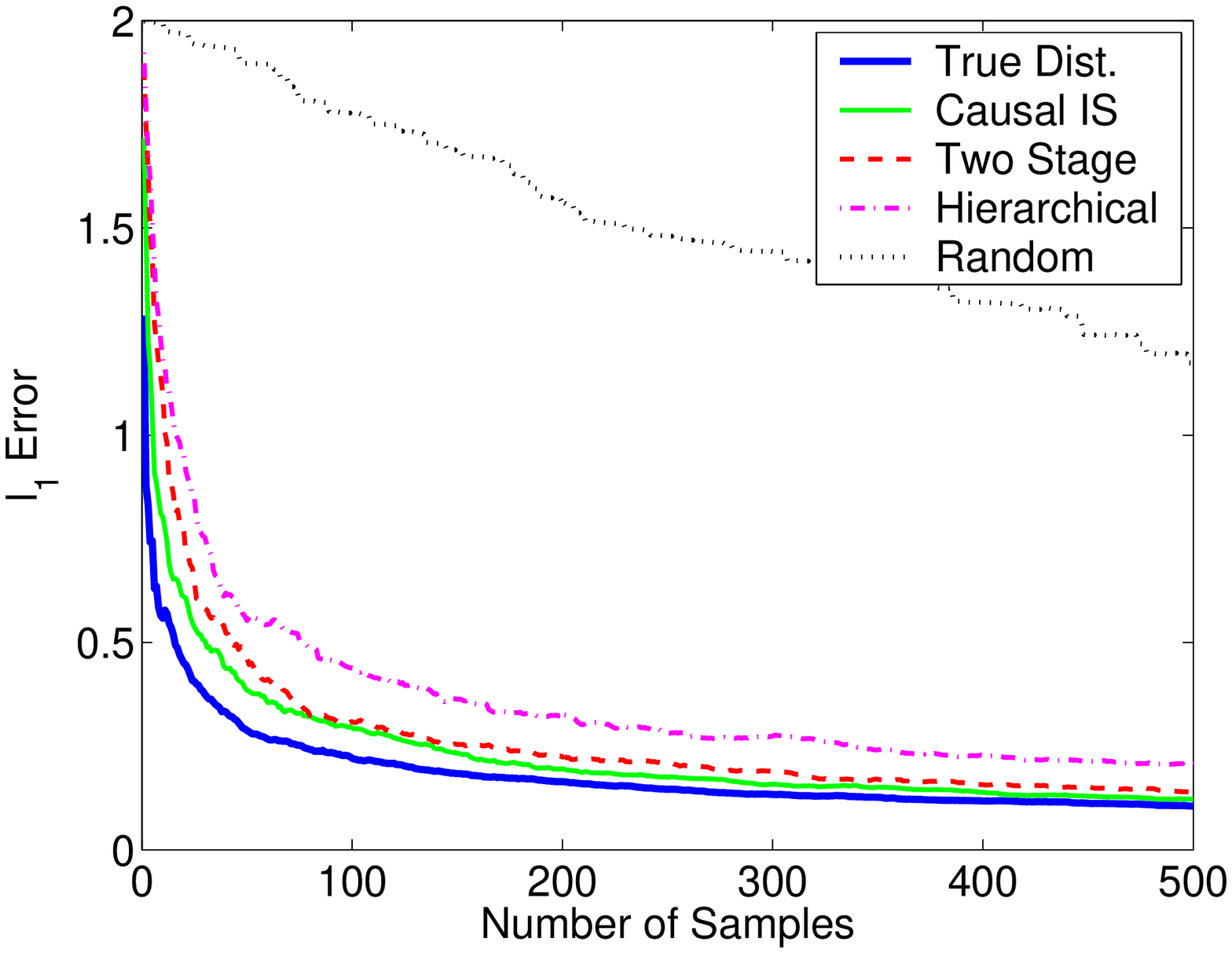}} %
\caption{$\ell_1$ error as a function of the number of importance
samples drawn for various sampling schemes in the following
scenarios: (a) a roughly uniform distribution on the permutations,
(b) moderately peaked distribution, (c) highly concentrated
distribution.  The curves in each figure were calculated by
averaging over 50 Monte Carlo simulations.} \label{fig:tv}
\end{figure*}

Figure~\ref{fig:tv} depicts the $\ell_1$ error between the true
and IS-based distributions on permutations, as a function of the
number of samples.  The curve labelled ``True Dist'' corresponds
to sampling from the true distribution, shown as a reference,  is
only possible when we can enumerate all permutations.  The
``Causal IS'' curve corresponds to the scheme described in
Section~\ref{subsec:causalIS}. The ``Two Stage'' and
``Hierarchical'' curves depict the performance for the
hierarchical schemes mentioned in Section~\ref{subsec:twoStageIS}.
Finally, ``Random'' refers to an approach where we sample from a
uniform distribution on permutations,  shown as a baseline
comparison.  Each curve was generated by averaging over 50 Monte
Carlo simulations. These curves depict performance using
up to 500 samples for a path with 720 possible orderings, which is
actually quite a generous helping of samples. In experiments with
Internet data we have encountered paths of up to length 27, and
observed good reconstruction performance using as few as  $2000$
samples (notice that $27! \simeq 10^{28}$).
Thus, performance for very few samples is of great
interest. As expected, all of the sampling schemes give the same
performance when  distribution is roughly uniform.  However, as
the distribution becomes more concentrated,
there is a clear difference between the various
sampling schemes.  The uniform sampling scheme naturally performs
the worst on more concentrated distributions. Of the schemes that
are practical for long paths, these results indicate that the causal
sampling scheme performs the best.

In terms of computational complexity, the causal sampler is the
simplest and fastest, requiring only $O(N)$ operations per sample
($N$ is the path length).  The two-stage sampler requires
$O\big((N/2)!\big)$ operations per sample, while the fully
hierarchical has computational complexity $O(N^2 \log N)$
operations per sample.  The conclusion is that the causal sampling
method described above is simple to implement, fast, and it
empirically outperforms more computationally complex schemes.

\begin{figure}
\centering %
\includegraphics[width=2.5in]{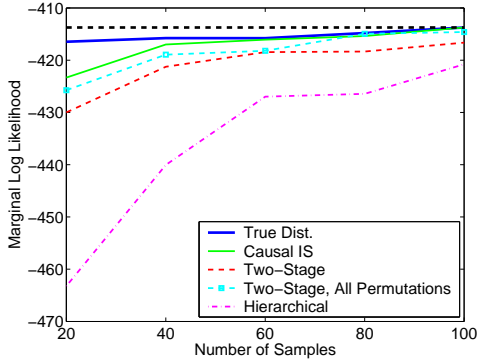} %
\caption{Using various approximate E-steps in the EM algorithm for
estimating the Markov transition matrix of a simulated network.
The horizontal dashed line at the top of the figure marks the
marginal log likelihood of the data using the transition matrix
derived using the correctly ordered paths.  The curves in this
figure correspond to the average over 10 Monte Carlo simulations.}
\label{fig:approxEcomp}
\end{figure}

To compare the efficacy of each sampling scheme within the EM
algorithm, we generated a random network with 250 nodes and
simulated 60 random sample paths through this network, ranging in
length from 4 to 10 hops. Then, we estimated a probability
transition matrix for the network using the EM algorithm with
different IS-based E-steps (with known endpoints for each path).
Figure~\ref{fig:approxEcomp} depicts the marginal log-likelihood
of the data, computed according to \eqref{eq:marginal} using the
probability transition matrices returned by the EM algorithm,  for
a number of samples-per-path between 20 and 100. The horizontal
dashed line at the top marks the marginal log likelihood computed
using a transition matrix estimated from correctly ordered paths.

\section{Monotonicity and Convergence}
\label{sec:convergence}

Well-known convergence results due to Wu and Boyles
\cite{wu83,boyles83} guarantee convergence of our EM algorithm
when the exact E-step is used. Let $\btheta^k = \big(\Amat^k,
\bpi^k\big)$ denote parameter estimates calculated at the $k$th EM
iteration using the exact EM expressions. By choosing
$\btheta^{k+1} = \big(\Amat^{k+1}, \bpi^{k+1}\big)$ according to
\eqref{eq:Mstep} in the M-step, our iterates satisfy the
\emph{monotonicity property}:
\begin{eqnarray}
  Q\left( \btheta^{k+1}; \btheta^k\right) &\ge&
  Q\left(\btheta^k; \btheta^k\right). \label{eq:mono}
\end{eqnarray}
The marginal log-likelihood \eqref{eq:marginal} is continuous in
its parameters $\Amat$ and $\bpi$ and it is bounded above.  In
this setting the monotonicity property \eqref{eq:mono} guarantees
that each exact EM update monotonically increases the marginal
log-likelihood, so the EM iterates converge to a local maximum.

When Monte Carlo methods are used in the E-step monotonicity is no
longer guaranteed since the M-step solves
\begin{eqnarray*}
  \widehat \btheta^{k+1} \equiv \big(\widehat{\Amat}^{k+1}, \widehat \bpi^{k+1}\big) &=&
  \arg \max_{\Amat, \bpi}
  \widehat Q\big(\Amat, \bpi; \Amat^k, \bpi^k\big),
\end{eqnarray*}
where $\widehat Q$ is defined analogously to $Q$ but with terms
$\bar \alpha_{t',t''}^{(m)}$ and $\bar r_{1,t'}^{(m)}$ replaced by
$\widehat \alpha_{t',t''}^{(m)}$ and $\widehat r_{1,t'}^{(m)}$,
their corresponding importance sampling approximations.
Consequently, care must be taken to ensure that $\widehat Q$
approximates $Q$ well enough so that the EM algorithm is not
swamped with error from the Monte Carlo estimates.

\begin{figure}
\centering %
\subfigure[][]{\includegraphics[width=2.75in]{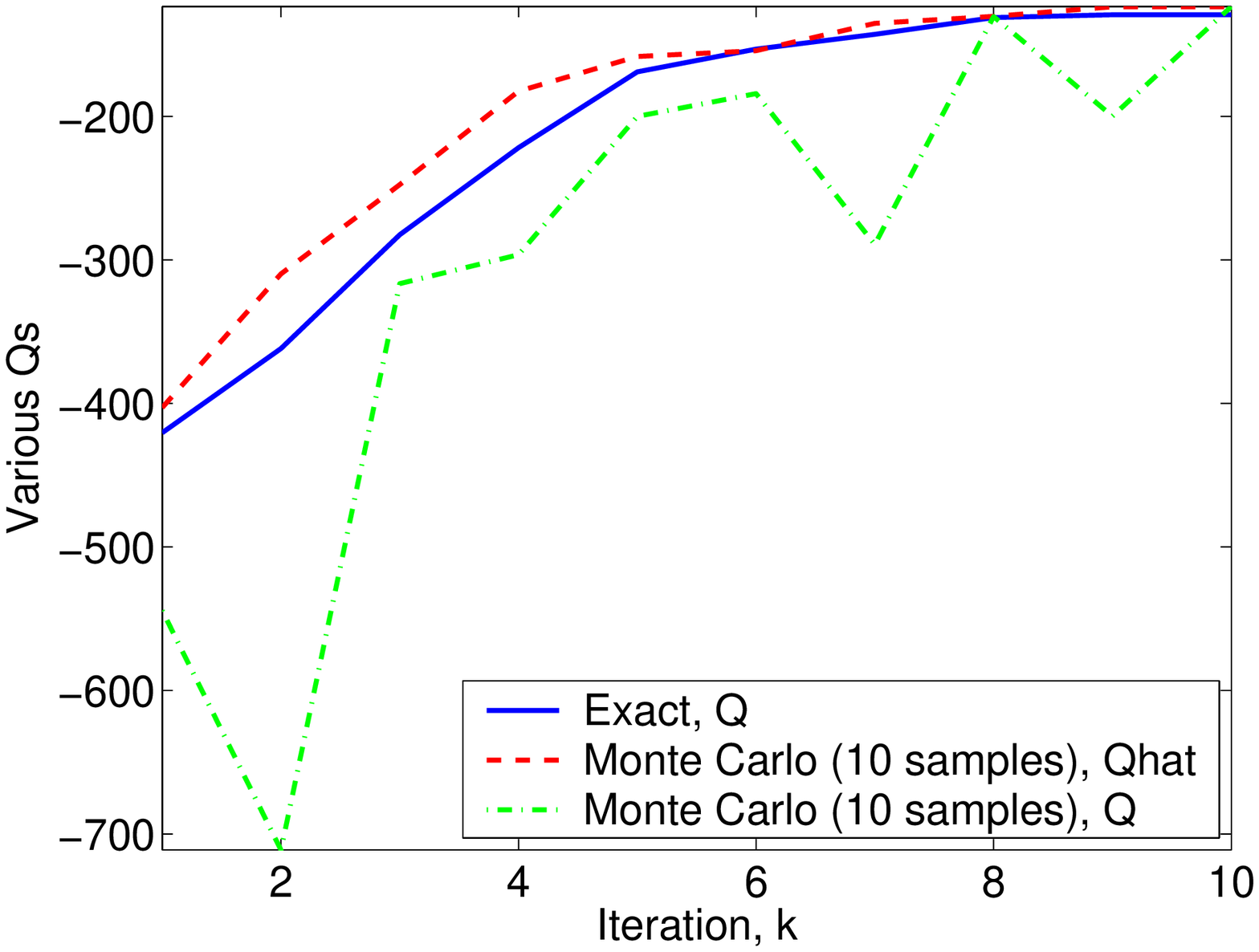}} %
\subfigure[][]{\includegraphics[width=2.75in]{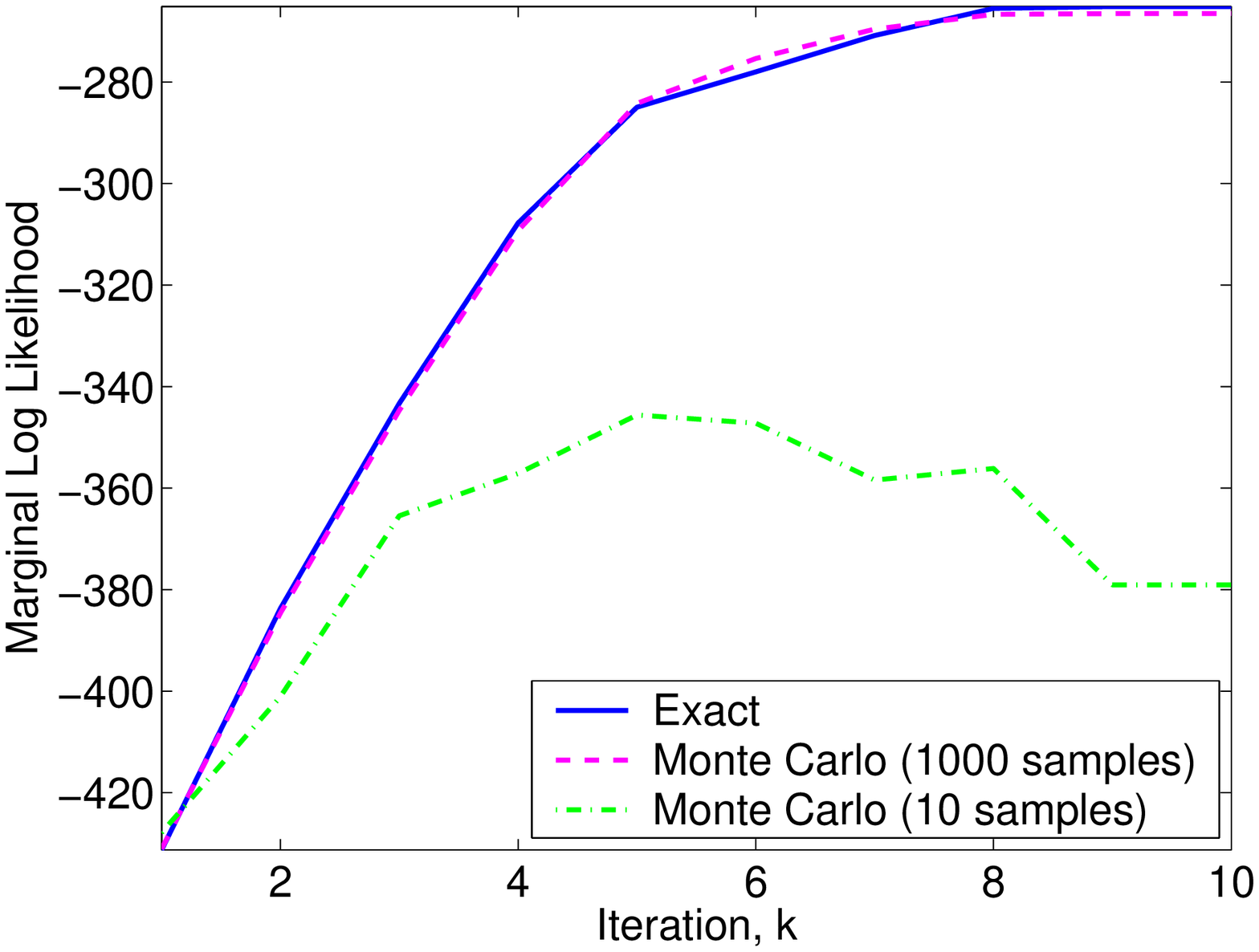}} %
\caption{An example with simulated observations illustrating that
the Monte Carlo EM algorithm may not result in monotonic increase
of the marginal log-likelihood if too few Monte Carlo samples are
used. The solid line in (a) is $Q(\btheta^{k+1}; \btheta^{k})$ for
exact EM iterations, the dashed line is $\widehat Q(\widehat
\btheta^{k+1}; \widehat \btheta^k)$ and the dash-dot line is
$Q(\widehat \btheta^{k+1}; \widehat \btheta^k)$ for Monte Carlo EM
iterations using only 10 samples. Even though $\widehat Q$
increases monotonically, $Q$ may not be monotonic for the Monte
Carlo EM algorithm.  Figure (b) depicts the marginal
log-likelihood for exact EM iterates and for two versions of the
Monte Carlo EM.  Monte Carlo EM performance closely resembles that
of the exact EM algorithm when sufficiently many importance
samples are used.} \label{fig:exactComp}
\end{figure}

To illustrate this issue, consider the following synthetic
example.  We generate 40 co-occurrence observations by taking a
random walk on a graph with 140 vertices.  Each co-occurrence has
between 4 and 8 vertices.  Figure~\ref{fig:exactComp}(a) plots
$Q(\btheta^k ; \btheta^{k-1})$ for the exact E-step, along with
$\widehat Q(\widehat \btheta^{k+1}; \widehat \btheta^k)$ and
$Q(\widehat \btheta^{k+1}; \widehat \btheta^k)$ for the
Monte Carlo EM algorithm using only 10 importance samples per
co-occurrence.  Although $\widehat Q(\widehat \btheta^{k+1};
\widehat \btheta^k)$ increases at each iteration, $Q(\widehat
\btheta^{k+1}; \widehat \btheta^k)$ clearly does not and the
monotonicity property does not hold.  This is apparent in
Figure~\ref{fig:exactComp}(b), where the dash-dot line shows the
progress of the marginal log-likelihood (our optimization
criterion) for the 10 sample Monte Carlo EM algorithm.  When
enough importance samples are used the Monte Carlo EM algorithm
performs comparably to the exact EM algorithm; see the dashed line
in Figure~\ref{fig:exactComp}(b) corresponding to a Monte Carlo EM
algorithm using 1000 importance samples per co-occurrence.  All
three instances of the EM algorithm used in this example start
from the same initialization.

Recently, researchers have considered the question of how many
importance samples should be used in a Monte Carlo E-step
\cite{booth01,caffo05,jank05}.  The goal is to balance
monotonicity and computational efficiency by using enough samples
to have a good chance at monotonicity while not using excessively
many samples.  Booth et al.~\cite{booth01} argue that if the same
number of importance samples is used at each EM iteration, then the
algorithm will eventually be swamped by Monte Carlo error and will
not converge. They also suggest requiring that a convergence
criterion be satisfied on multiple successive iterations since the
criterion may be met prematurely due to poor Monte Carlo
approximations.

Caffo et al.~\cite{caffo05} propose a method for automatically
adapting the number of Monte Carlo samples used at each EM
iteration.  To lighten notation, we drop the superscripts $k$ and
$k+1$.  Let $\Delta(\btheta) = Q(\btheta; \btheta') - Q(\btheta';
\btheta')$ and $\widehat \Delta(\btheta) = \widehat Q(\btheta;
\btheta') - \widehat Q(\btheta'; \btheta')$. Furthermore, let
$\widehat \btheta = \arg \max_{\btheta} \widehat Q(\btheta;
\btheta')$, where $\btheta' = \btheta^k$ was determined
at the previous EM iteration.  Recall that $L$
importance samples are used to calculate $\widehat Q$.  The
algorithm of Caffo et al.~is based on a Central Limit Theorem-like
approximation in which they show that $\sqrt{L} \big(\widehat
\Delta(\widehat \btheta) - \Delta(\widehat \btheta)\big)$
converges in distribution to the standard normal.  Observe that
the monotonicity property \eqref{eq:mono} is equivalent to the
condition $\Delta(\widehat \btheta) \ge 0$.  Although
$\Delta(\widehat \btheta)$ cannot be computed without computing
the exact sufficient statistics $\{\bar \alpha_{t',t''}^{(m)}\}$
and $\{\bar r_{1,t'}^{(m)}\}$, we can compute $\widehat
\Delta(\widehat \btheta)$.  Their scheme then amounts to
increasing the number of Monte Carlo samples until $\widehat
\Delta(\widehat \btheta) > \epsilon$ for a user-specified
$\epsilon > 0$.  Then, applying an asymptotic standard normal tail
approximation, they obtain a statement of the form $\Pr\left(
\widehat\Delta(\widehat \btheta) - \Delta(\widehat \btheta) \ge
\epsilon\right) \le \delta(\epsilon)$. Based on this statement
they claim that monotonicity holds with probability at least $1 -
\delta(\epsilon)$.  They further remark that if a different
$\epsilon_k$ is chosen at each iteration, so that
$\sum_{k=1}^{\infty} \delta(\epsilon_k) < \infty$, then, by the
Borel-Cantelli Lemma, $\Pr\left(\widehat\Delta(\widehat \btheta) -
\Delta(\widehat \btheta) \ge \epsilon_k \ \mbox{i.o.}\right) = 0$,
so there exists a $K > 0$ such that $\widehat \Delta(\widehat
\btheta) - \Delta(\widehat \btheta) < \epsilon_k$ for all $k \ge
K$ with probability 1; \emph{i.e.}, eventually every EM update is
monotonic.  Of course, in practice, the algorithm is terminated
after a finite number of iterations, so we may never reach the
stage where all iterates are monotonic.

Notice that for the monotonicity condition $\Delta(\widehat
\btheta) \ge 0$ to truly hold in the above framework, the events
\begin{eqnarray*}
  \left\{\widehat\Delta(\widehat \btheta) -
  \Delta(\widehat \btheta) \le \epsilon\right\} & \mbox{
  and } & \left\{ \widehat \Delta(\widehat \btheta) \ge \epsilon
  \right\}
\end{eqnarray*}
must occur simultaneously.  Because the probabilistic bound above
only addresses one of these events we refer to this type of result
as guaranteeing an $(\epsilon, \delta)$-\emph{probably
approximately monotonic} update, or PAM for short.  More
generally, an $(\epsilon,\delta)$-PAM result states that with
probability at least $1-\delta$, the update will be
$\epsilon$-approximately monotonic; \emph{i.e.}, $\widehat
\Delta(\widehat \btheta) - \Delta(\widehat \btheta) \le \epsilon$
implies $\Delta(\widehat \btheta) \ge -\epsilon$, because,
by definition, $\widehat \Delta(\widehat \btheta)\geq 0$.

Rather than resorting to asymptotic approximations to obtain such
a result, we can take advantage of the specific form of $Q$ in our
problem to obtain the finite-sample PAM result next presented. Recall
that independent importance samples are drawn for each
co-occurrence observation in the Monte Carlo E-step. Denote by
$L_m$ the number of importance samples used to compute sufficient
statistics for observation $\bx^{(m)}$.  Exact E-step computation
for this observation requires $N_m!$ operations.  Similarly, we
should expect that larger observations will require more
importance samples for two reasons: 1) there are more sufficient
statistics associated with this observation ($N_m^2$ in total),
and 2) there are more ways to shuffle these observations.

In the previous section we derived closed form expressions for the
importance sample weights, $z_i = P[\br | \bx, \Amat,\bpi]/
R[\br | \bx, \Amat, \bpi]$, where $P$ is the target
distribution and $R$ is the importance sampling distribution.  A
key assumption was made that $P$ is absolutely continuous with
respect to $R$; that is, $P[\br | \bx, \Amat, \bpi] = 0$ for every
permutation $\br$ with $R[\br | \bx, \Amat, \bpi] = 0$.  We adopt
the convention $0/0 = 0$ so that $z_i = 0$ for such samples.  This
guarantees that $z_i < \infty$.  The bounds below depend on the
quality of our importance sample estimators as gauged by
\begin{eqnarray}
  b_m &=& \max_{\br \in \Psi_{N_m}} \frac{P[\br | \bx^{(m)},
  \Amat, \bpi]}{R[\br | \bx^{(m)}, \Amat, \bpi]}.
  \label{eq:mc_bm}
\end{eqnarray}
Because the set $\Psi_{N_m}$ is finite, $P$ and $R$ have finite
support, and the maximum is well-defined.  If $R$ matches the
target distribution $P$ well then $b_m$ should not be very large.

There is one other subtlety we will account for in our bounds.
Because $\widehat Q(\btheta; \btheta')$ has terms $\log A_{i,j}$
and $\log \pi_i$, in practice we typically bound $A_{i,j}$ and
$\pi_i$ away from zero to ensure that $\widehat Q$ does not go to
$-\infty$.  To this end, we will assume that $\widehat A_{i,j} \ge
\thetamin$ and $\widehat \pi_i \ge \thetamin$ for some $0 <
\thetamin < |S|^{-1}$.  The upper bound on $\thetamin$ ensures it
is still possible to satisfy the constraints
\eqref{stochasticmatrix}. Generally we choose $\thetamin$ very
close to zero; at machine precision, for example.

We have the following finite-sample PAM result for our Monte Carlo
EM algorithm.

\begin{thm} \label{thm:pam}
Let $\epsilon > 0$ and $\delta > 0$ be given and assume there
exists $\thetamin \in (0, |S|^{-1})$ such that $A'_{i,j} \ge
\thetamin$ and $\pi'_i \ge \thetamin$ for all $i$ and $j$. If
\begin{eqnarray}
  L_m &=& \frac{2 T^2 N_m^4 b_m^2 \ |\log \thetamin|^2}{\epsilon^2}
  \log\left(\frac{2 N_m^2}{1 - (1-\delta)^{1/T}}\right)
  \label{eq:Lm_pam}
\end{eqnarray}
importance samples are used for the $m$th observation then
$\widehat \Delta(\widehat \btheta) - \Delta(\widehat \btheta) <
\epsilon$ with probability greater than $1 - \delta$.
\end{thm}

The proof of Theorem~\ref{thm:pam} appears in
Appendix~\ref{sec:pamproof}.  Because $\widehat \Delta(\widehat
\btheta) \ge 0$ by definition, the theorem guarantees that
$\Delta(\widehat \btheta) > -\epsilon$ with probability greater
than $1-\delta$.

Recall that exact E-step computation requires $N_m!$ operations
for the $m$th observation.  The bound above stipulates that the
number of importance samples required is proportional to $N_m^4
\log N_m^2$, and generating one importance sample requires $N_m$
operations.  Thus, the computational complexity of a PAM Monte
Carlo update only depends on $N_m^5 \log N_m^2$, which clearly
demonstrates that the computational complexity of the Monte Carlo
E-step depends polynomially on the $N_m$ in comparison
to exponential dependence for the exact E-step.

To put this result in perspective, observe that the value of $L_m$
given by \eqref{eq:Lm_pam} is roughly a factor of $T$ away
from the value we would expect based on an asymptotic variance
calculation.  Ignoring constants and log terms, for fixed
$\btheta$ we have
\begin{eqnarray*}
  \var\left(\widehat \Delta(\btheta)\right) &\approx&
  \var\left(\sum_{m=1}^T \sum_{t',t''=1}^{N_m} \widehat
  \alpha_{t',t''}^{(m)} + \sum_{m=1}^T \sum_{t'=1}^{N_m} \widehat r_{1,t'}^{(m)}\right)\\
  &=& \sum_{m=1}^T \var\left(\sum_{t',t''=1}^{N_m} \widehat
  \alpha_{t',t''}^{(m)} + \sum_{t'=1}^{N_m} \widehat r_{1,t'}^{(m)}\right),
\end{eqnarray*}
since independent sets of importance samples are used to calculate
sufficient statistics for different observations.  It is easily
shown that the variance of an individual approximate statistic
$\widehat \alpha_{t',t''}^{(m)}$ or $\widehat r_{1,t'}^{(m)}$
decays according to the parametric rate; \emph{i.e.},
$\var(\widehat \alpha_{t',t''}^{(m)}) \simeq 1/L_m$.  In total,
there are $N_m^2$ sufficient statistics for the $m$th observation,
and they are all potentially correlated since they are functions
of the same set of importance samples. Then we have
\begin{eqnarray*}
  \var\left(\widehat \Delta(\btheta)\right) &\approx& \sum_{m=1}^T
  \frac{\left(N_m^2\right)^2}{L_m}.
\end{eqnarray*}
To drive $\var\left(\widehat \Delta(\btheta)\right)$ down to a
constant level, independent of $T$ and $N_m$, we need $L_m \propto
T N_m^4$.  The additional factor of $T$ in our bound is
essentially an artifact from the union bound.

Although the PAM result is pleasing, we would prefer to guarantee
\emph{monotonicity} with high probability, not just
\emph{approximate} monotonicity.  Let $\btheta^* = \arg
\max_{\btheta} Q(\btheta ; \btheta')$.  By bounding
$\Delta(\widehat \btheta) - \Delta(\btheta^*)$ instead of
$\widehat \Delta(\widehat \btheta) - \Delta(\widehat \btheta)$ we
obtain the following stronger result guaranteeing monotonicity
with high probability. Instead of restricting $A_{i,j}\ge
\thetamin$ and $\pi_i \ge \thetamin$, we need to assume the
variables $\bar \alpha_{t',t''}^{(m)}$ and $\bar r_{1,t'}^{(m)}$
are bounded away from zero.  This is stronger than the previous
assumption in the sense that it implies $A_{i,j}$ and $\pi_i$ are
bounded away from zero.

\begin{thm} \label{thm:probmono1}
Let $\delta > 0$ be given and assume there exists $\lambda > 0$
such that $\bar \alpha_{t',t''}^{(m)} > \lambda$ and $\bar
r_{1,t'}^{(m)} > \lambda$, for all $t'$ and $t''$.  If
\begin{eqnarray}
L_m &=& \frac{27 b_m}{\lambda} \left(\frac{2 \sum_{m=1}^T N_m +
\Delta(\btheta^*)}{\Delta(\btheta^*)}\right)^2 \log\left(\frac{4
\sum_{m=1}^T N_m^2}{\delta}\right) \label{eq:Lm_pm}
\end{eqnarray}
importance samples are used for the $m$th observation, then
$\Delta(\widehat \btheta) \ge 0$ with probability at least $1 -
\delta$.
\end{thm}

The proof of Theorem~\ref{thm:probmono1} appears in
Appendix~\ref{sec:probmonoproof}.  Similar to the PAM bound given
in Theorem~\ref{thm:probmono1}, the computational complexity
required for a \emph{probably monotonic} update also depends
polynomially on $N_m$.

Note that $L_m$ depends on $\Delta(\btheta^*)$.  By definition,
$\Delta(\btheta^*) \ge 0$ at every iteration, and typically
$\Delta(\btheta^*)$ is large at earlier EM iterations and
approaches zero as the algorithm converges.  This dependence
supports the observation of Booth et al.~mentioned earlier, that
the number of importance samples ought to increase at each
iteration.

The main assumption of Theorem~\ref{thm:probmono1} is that the
sufficient statistics are bounded away from zero at each
iteration.  We motivate this assumption by observing that if the
algorithm is initialized with $\pi_i^0 > 0$ and $A_{i,j}^0 > 0$
for all $i,j$ then, recalling the closed form E-step expressions,
the sufficient statistics do not vanish in a finite number of EM
iterations.


Note that if we use different $\delta_k$ at each EM iteration,
chosen such that $\sum_{k=1}^\infty \delta_k < \infty$, then by
the Borel-Cantelli Lemma one can argue that
$\Pr\big(\Delta(\widehat \btheta) < 0 \ \mbox{ i.o.}\big) = 0$. In
other words, eventually all EM iterates result in a monotonic
increase of the marginal log-likelihood.

In addition to demonstrating that the Monte Carlo EM algorithm has
polynomial computational complexity, these bounds give a useful
guideline for determining how many importance samples should be
used.  However, because they involve worst-case analysis, the
numbers of samples dictated by these bounds tend to be on the
conservative side.  For example, in the Internet experiments
described in Section~\ref{sec:sims}, $T = 249$ and $\bar N = 17$.
Theorem~\ref{thm:probmono1} suggests that roughly 72 million
importance samples should be used per observation.  However, in
our experiments we find that the algorithm exhibits reasonable
performance on this data set using as few as $2,000$ samples per
observation.  Of course, in practice, we do not have direct access
to the parameters $b_m$, $\lambda$, or $\Delta(\btheta^*)$, so
these bounds cannot be used as explicit guidelines.

\section{Experimental Results}
\label{sec:sims}

In this section, we evaluate the performance of our \emph{network
inference from co-occurrences} (NICO) algorithm on simulated data
and on data gathered from the public Internet.  In the results
reported below, network reconstructions are obtained by first
estimating an initial state distribution and probability
transition matrix via the EM algorithm.  Then, we compute the most
likely order of each observation according to the inferred model
and use this ordering to reconstruct a feasible network.  The EM
algorithm cannot be guaranteed to converge to a global maximum
(the marginal log-likelihood is not concave) and there may even
be multiple global maxima.  To address this issue, we rerun the
EM algorithm from multiple random initializations and report
the collective results.

We compare the performance of our algorithm with that of the
\emph{frequency method} (FM), defined in \cite{rabbat05} and
mentioned in Section \ref{introduction}.  The FM also reconstructs a network
topology by estimating an order of the vertices in each
observation.  This method individually determines each path
ordering independently by sorting the elements in the path
according to a score computed from pair-wise co-occurrence
frequencies involving the source and destination of the path.  It
is possible that multiple vertices may receive identical FM
scores, in which case their sorting would be arbitrary (one
could exchange elements with identical scores without violating
the FM criteria). In fact, we observe this phenomenon in many of
our experiments.  Ties are resolved by choosing a random order for
elements with identical scores.  Multiple restarts are also
performed using the FM, yielding a collection of feasible
solutions.

The quality of a network reconstruction is determined by a
quantity we term the \emph{edge symmetric difference} error.
Because the nodes in the network have unique labels, the goal of
any reconstruction scheme is to determine which vertices are
connected by an edge.  The edge symmetric difference error is
defined as the sum of the number of false positives (edges
appearing in the reconstructed network which do not exist in the
true network) and the number of false negatives (edges in the true
network not appearing in the reconstructed network).

\subsection{Simulated Networks}

Our synthetic data is obtained as described next. A
network is generated according to a random geometric graph model:
50 vertices are thrown at random in the unit square, and two
vertices are connected with an edge if the Euclidean distance
between them is less than or equal to $\sqrt{\log(50)/50}$.  This
threshold guarantees that the graph is connected with high
probability.  Groups of nodes are randomly chosen as sources and
destinations, transmission paths are generated between each
source-destination pair according to either a shortest path or
random routing model, and then co-occurrence observations are
formed from each path.  We keep the number of sources fixed at 5
and vary the number of destinations between 5 and 40, to see how
the number of observations effects performance.  Each experiment
is repeated on 100 different topologies, using 10 restarts of both
NICO and the FM per configuration.  Exact E-step calculation is
used for observations with $N_m \le 12$, and causal importance
sampling (2000 samples) is used for longer paths. The longest
observation in our data was obtained by random routing and
has $N_m = 19$ (notice that $19!\simeq 10^{17}$).

\begin{figure}
\centering %
\subfigure[Shortest path
routes]{\includegraphics[width=2.5in]{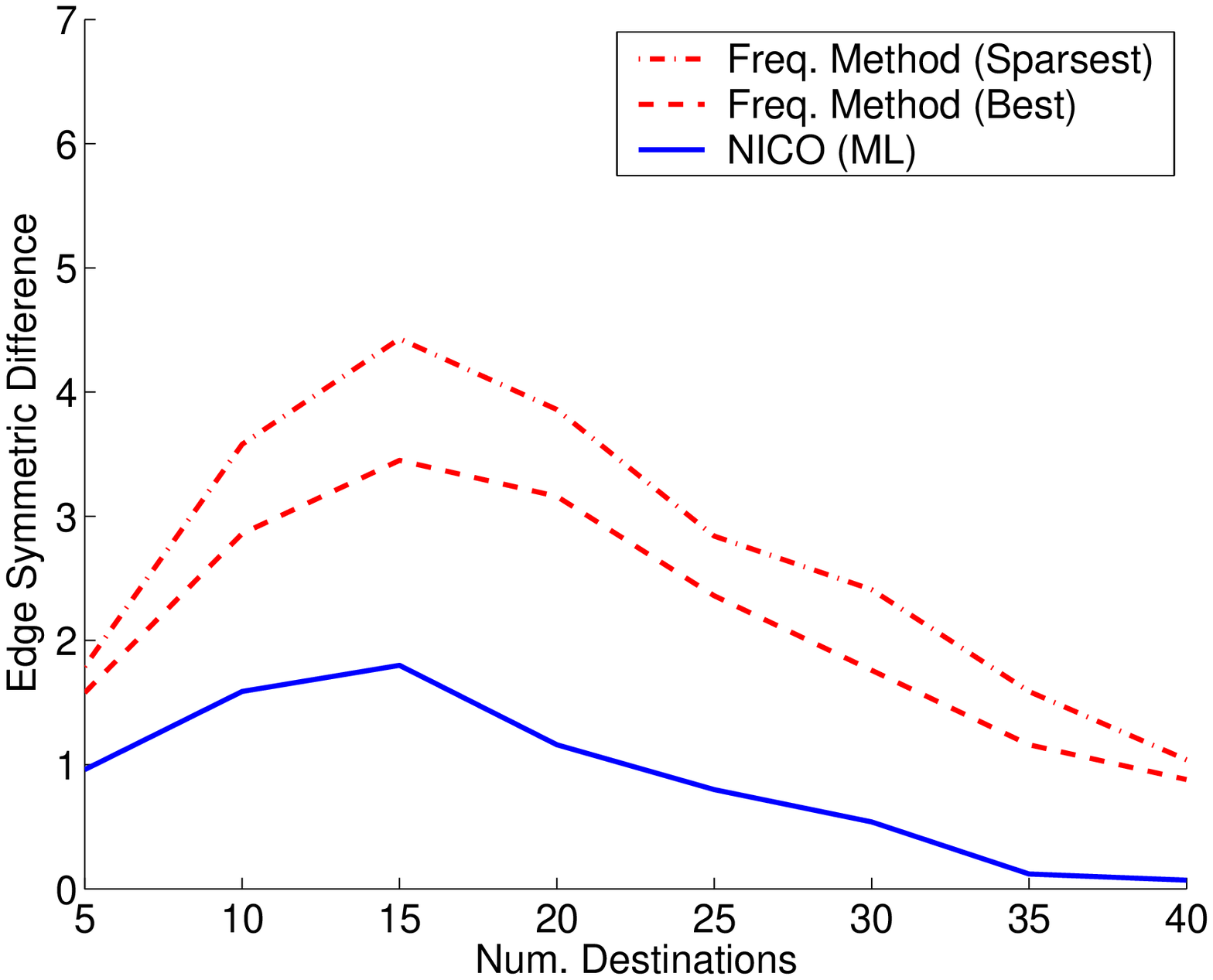}} %
\hspace{.5in} %
\subfigure[Random
routes]{\includegraphics[width=2.5in]{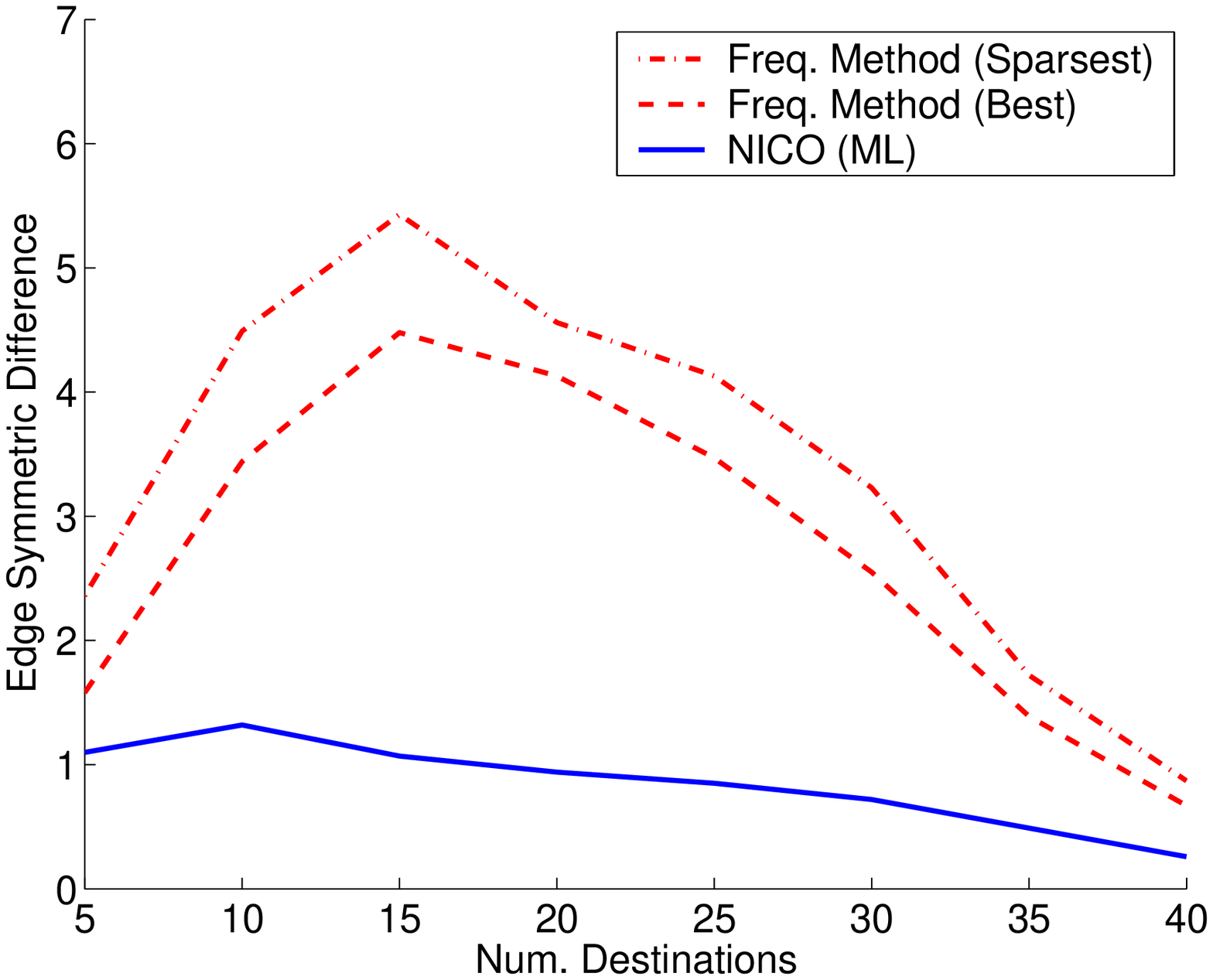}} %
\caption{Edge symmetric differences between inferred networks and
the network one would obtain using co-occurrence measurements
arranged in the correct order.  Performance is averaged over 100
different network configurations.  For each configuration 10 NICO
and FM solutions are obtained via different initializations.  We
then choose the NICO solution yielding the largest likelihood, and
compare with both the sparsest and clairvoyant best FM solution.}
\label{fig:synthData}
\end{figure}

Figure~\ref{fig:synthData} plots edge symmetric difference
performance for synthetic data generated using (a) shortest path
routing and (b) random routing.  The edge symmetric difference
error is computed between the inferred network and the graph
obtained from correctly ordered observations.  Of the 10 solutions
corresponding to different NICO initializations, we use the one
based on parameter estimates yielding the highest likelihood
score.  For this simulation, the most likely NICO solution also
always resulted in the best edge symmetric difference error.

The FM does not provide a similar mechanism for ranking different
solutions. A possible heuristic would be to choose the sparsest
solution (with fewest edges).  The figures show performance
for both this heuristic, and clairvoyantly choosing
the best (lowest error) solution of the 10. In fact, using the
sparsest solution does better than just choosing a FM solution at
random but not as well as using the clairvoyant best. In these
simulations, NICO consistently outperform the FM.

Notice that both algorithms exhibit their worst performance at an
intermediate number of destinations.  When very few destinations
are used the measured topology closely resembles a tree,
regardless of the underlying routing mechanism. Relative
frequencies of co-occurrence accurately reflect the network
distance of each internal vertex from the path endpoints.  At the
other extreme, when many destinations are used, there is
significant overlap among the co-occurrence observations which
aids in localizing vertices.  In general, the FM seems to be much
more sensitive to the amount of data available.

As expected, the FM generally performs better on shortest path
data than it does on random routes.  When routes are generated
randomly the corresponding topology is less tree-like and
pair-wise co-occurrence frequencies do not reflect network
distances.  Because NICO is not based on a particular routing
paradigm it performs similarly in both scenarios, possibly even a
little better when routing is random.

\subsection{Internet Data}

We have also studied the performance of our algorithm on
co-occurrence observations gathered from the Internet.  Using
\texttt{traceroute} we have collected data describing roughly 250
router-level paths from sources at the University of
Wisconsin-Madison, the \emph{Instituto Superior T\'ecnico} in
Lisbon, and Rice University to 83 servers affiliated with
corporations, universities, and governments around the world.  Our
motivation for using this type of data is two-fold.  First,
\texttt{traceroute} allows us to measure the true order of
elements in each path so that we have a ground truth to validate
our results against. Second, and more importantly, the data comes
from a real network where, presumably, paths are not generated
according to a first-order Markov model. This allows us to gauge
the robustness of the proposed model and to evaluate how well it
generalizes to realistic scenarios.  The ground truth network
contains a total of 1105 nodes and 1317 edges, and the longest
observed path has length 27.

For this data set we rerun FM and NICO each from 50 random
initializations and look at performance across all solutions
rather than focusing on the maximum likelihood or clairvoyant
best.  The exact E-step is used to compute $\bar \alpha$ for paths
of up to 9 hops. For paths longer than 9 hops, we use the causal
importance sampling described in Section~\ref{subsec:causalIS},
with 2000 samples per observation.

\begin{figure}
\centering %
\includegraphics[width=2.5in]{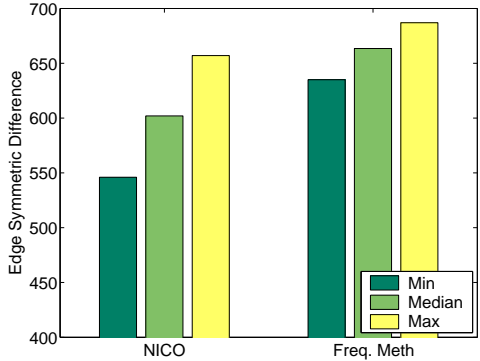}
\caption{Edge symmetric difference error comparison of NICO and FM
on Internet data. The reported values come from 50 random
initializations of each algorithm.} \label{fig:tracertSymmDiff}
\end{figure}

Minimum, median, and maximum edge symmetric difference errors are
shown in Figure~\ref{fig:tracertSymmDiff}.  Both algorithms have
seemingly high error rates, as there are roughly 1300 links in the
true network.  However keep in mind that both algorithms are
attempting to fill in the entries of a roughly $1100 \times 1100$
matrix. For 50 networks constructed by choosing a random order for
the elements of each observation the average edge symmetric
difference error was 4300, so both algorithms are indeed doing
considerably better than random guessing.  NICO performance is
again noticeably better than that of the FM; the NICO average
error is better than that of the best FM reconstruction, and the
worst case NICO reconstruction is on par with the average FM
performance.  We also note that the number of false positives and
false negatives in a reconstruction using either scheme tend to be
roughly equal (each constituting half of the edge symmetric
difference error).

\begin{figure}
\centering %
\includegraphics[width=2.5in]{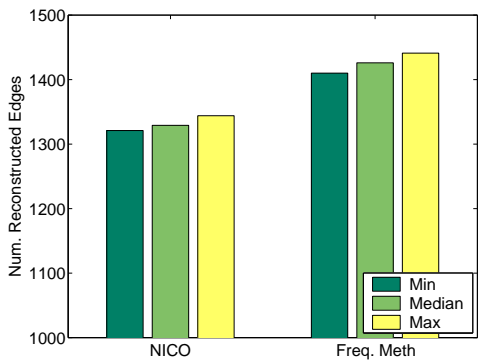}
\caption{Number of edges in networks reconstructed using each
method. The median number of edges per reconstruction is 1329 for
NICO and 1426 for FM.  The true network has 1317 edges, and so it
appears that NICO does a better job of capturing the complexity of
the true network.} \label{fig:tracertNumEdges}
\end{figure}

Figure~\ref{fig:tracertNumEdges} shows statistics for the number
of edges in the reconstructed networks.  There is an interesting
correlation between the number of edges and reconstruction
accuracy in this example.  As seen above, the typical NICO
reconstruction is more accurate, in terms of edge errors, than a
FM reconstruction.  NICO also consistently returns a sparser
estimate.  The median number of links in a NICO reconstruction is
1329, whereas the median number of links in a FM reconstruction is
1426.  There are 1317 edges in the true network, so in this sense
the NICO reconstructions more accurately reflect the inherent
level of complexity in the true network.

\begin{figure}
\centering %
\includegraphics[width=2.5in]{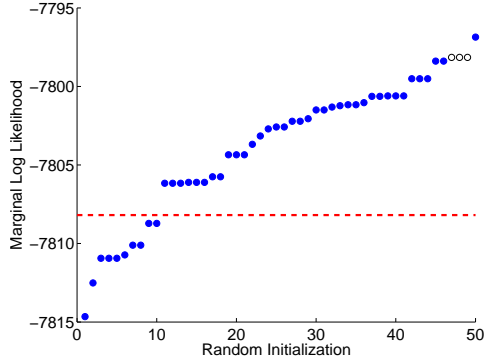}
\caption{Marginal log likelihood values for different random
initializations of NICO, sorted in ascending order. The three
hollow circles correspond to the solutions which achieve the
lowest edge symmetric difference error of all NICO trials. The red
line shows the marginal log likelihood value computed using the
true path orders to estimate a Markov transition matrix.  Most
NICO solutions have higher marginal log-likelihood than the true
topology, suggesting that our generative model does not accurately
describe Internet topology data.} \label{fig:tracertLogLik}
\end{figure}

Marginal log-likelihood values for each of the 50 NICO estimates
are depicted in Figure~\ref{fig:tracertLogLik}. The marginal
log-likelihood, given by \eqref{eq:marginal}, is the cost function
being optimized by the EM algorithm.  In contrast to the
experiments with simulated data reported above, there is no
exact correlation between higher marginal likelihood values and
lower edge symmetric difference error for this example.  The
topology with the highest likelihood value results in an edge
symmetric difference error of 627.  This is better than the
clairvoyant best FM error, but only average for NICO. The three
repetitions which returned a topology with the lowest symmetric
difference error had the next highest likelihood value, as
indicated by the three hollow circles in the figure.  The dashed
line shows the likelihood value based on a transition matrix
estimated using the true path orders as measured by
\texttt{traceroute}.  Notice that the majority of the NICO
solutions have a higher marginal likelihood than the true
topology.  This suggests that our generative model may not be the
best match for Internet topology data.  Still the overall
performance of our algorithm is encouraging.

\section{Discussion and Ongoing Research}
\label{sec:conc}

This paper presents a novel approach to network inference from
co-occurrence observations.  A co-occurrence observation reflects
which vertices are activated by a particular transmission through
the network, but not the order in which they are activated.  We
model transmission paths as a random walks on the underlying graph
structure.  Co-occurrence observations are modelled as i.i.d.
samples of the random walk subjected to a random permutation which
accounts for the lack of observed path order.  Treating the random
permutations as latent variables we derive an
\emph{expectation-maximization} (EM) algorithm for efficiently
computing maximum likelihood or maximum \emph{a posteriori}
estimates of the random walk parameters (initial state
distribution and transition matrix).

The complexity of the EM algorithm is dominated by the E-step
calculation which is exponential in the length of the longest
transmission path.  In order to handle large networks, we describe
fast approximation methods based on importance sampling and Monte
Carlo techniques.  We derive concentration-style bounds on the
accuracy of the Monte Carlo approximation.  These bounds prescribe how
many importance samples must be used to ensure a monotonic increase in
the log-likelihood, thereby guaranteeing convergence of the algorithm
with high probability.  The resulting Monte Carlo EM computational
complexity only depends polynomially on the length of the longest
path.

To obtain a network reconstruction, we determine the most likely
order for each co-occurrence observation according to the Markov
chain parameter estimates, and then insert edges in the graph
based on these ordered transmission paths.  This procedure always
produces a feasible reconstruction.  The parameter estimates
produced by the EM algorithm may be useful for other tasks such as
guiding an expert to alternative reconstructions by assigning
likelihoods to different permutations, or predicting unobserved
paths through the network as in \cite{justice05}.  One could also
analyze properties of an ensemble of solutions obtained by running
the EM algorithm from different initializations, and then posit a
new set of experiments to be conducted based on this analysis.

The transition matrix parameter $A_{i,j}$ can be interpreted as
estimates of the probability that a transmission will be passed
from vertex $i$ to $j$, conditioned on the path reaching $i$; that
is, $A_{i,j} = P[Z_{k+1} = j | Z_{k} = i]$.  In particular, they
\emph{are not} estimates of the probability of a link existing
from $i$ to $j$.  Since $\Amat$ is a stochastic matrix, each row
must sum to 1, and so if vertex $i$ is connected to many other
nodes then the unit mass is being spread over more entries. We can
obtain joint probabilities, $P[Z_{k} = i, Z_{k+1} = j]$, via Bayes
theorem,
\begin{eqnarray*}
P[Z_{k+1} = j | Z_{k} = i] = \frac{P[Z_{k} = i, Z_{k+1} =
j]}{P[Z_{k} = i]},
\end{eqnarray*}
where $P[Z_{k}=i]$ is the stationary distribution of the chain
(not necessarily equal to the initial state distribution).  These
joint probabilities (appropriately scaled versions of the
transition matrix entries) more accurately reflect the likelihood
of there being an edge from $i$ to $j$, based on our estimates.

Our future work involves extending and generalizing both algorithmic
and theoretical aspects of this work.  In our experiments we found
that our current model leads to reasonable Internet reconstructions,
but we feel there is room for improvement.  For example, the structure
of Internet paths may depend strongly on the destination of the
traffic.  We are currently investigating models based on ``mixtures of
random walks'' to account for this added level of dependency.

Co-occurrence observations naturally arise from transmission
\emph{paths} in communication network applications and, to a
degree, in biological, social, and brain networks as well.
However the physical mechanisms driving interactions in the latter
three applications may also correspond to more general connected
subgraph structures such as trees or directed acyclic graphs.
Extending our methods in this fashion is easily accomplished in
theory, however the computational complexity may be significantly
increased when more general structures are considered.

In this paper we have also restricted our attention to noise-free
observations.  We are also interested in extending our algorithm to
handle the case where observations reflect a soft probability that a
given vertex occurred in the path rather than hard, ``active" or ``not
active", binary observations.  This extension is relevant in many
applications including the inference of signal transduction networks
(in systems biology) where co-occurrence observations are themselves
the result of inference procedures run on experimental data.

\appendix

\section{Proof of Theorem~\ref{thm:pam}} \label{sec:pamproof}

There are two main steps in the proof of Theorem~\ref{thm:pam}.
First, we derive a concentration inequality for the importance
sample approximations, $\widehat \alpha_{t',t''}^{(m)}$ and
$\widehat r_{1,t'}^{(m)}$.  Then we use the inequality to
construct a bound for $\widehat \Delta(\widehat \btheta) -
\Delta(\btheta)$.

Recall the expressions \eqref{eqn:is_r} and \eqref{eqn:is_alpha}
for importance sample approximations calculated in the Monte Carlo
E-step.  Both are of the general form $\widehat \mu_L =
\frac{\sum_{i=1}^L Z(\br^i) X(\br^i)}{\sum_{i=1}^L Z(\br^i)}$,
where $Z : \Psi_N \rightarrow [0,b]$ and $X : \Psi_N \rightarrow
\{0,1\}$, and they are approximating $\mu = \sum_{\br \in \Psi_N}
X(\br) P[\br | \bx, \Amat, \bpi]$.  Note that $\E[\widehat \mu_L]
\ne \mu$, so standard concentration results such as Hoeffding's
inequality or McDiarmid's bounded-differences inequality do not
directly apply; \emph{e.g.}, consider the case $L=1$:
\begin{eqnarray}
  \E\left[\frac{Z(\br^1) X(\br^1)}{Z(\br^1)}\right] &=& \sum_{\br \in \Psi_N}
  X(\br) R[\br | \bx, \Amat, \bpi] \\
  &\ne& \sum_{\br \in \Psi_N} X(\br) P[\br | \bx, \Amat, \bpi].
\end{eqnarray}
We can, however, show that $\widehat \mu_L$ yields an
asymptotically consistent estimate of $\mu$.  Observe that
\begin{eqnarray}
  \E[Z(\br^i)] &=& \sum_{\br \in \Psi_N} \frac{P[\br | \bx, \Amat,
  \bpi]}{R[\br | \bx, \Amat, \bpi]} R[\br | \bx, \Amat, \bpi] \\
  &=& 1,
\end{eqnarray}
since $P$ is a probability distribution on $\Psi_N$, and
\begin{eqnarray}
  \E[Z(\br^i) X(\br^i)] &=& \sum_{\br \in \Psi_N} \frac{P[\br | \bx, \Amat,
  \bpi]}{R[\br | \bx, \Amat, \bpi]} X(\br) R[\br | \bx, \Amat,
  \bpi]\\
  &=& \sum_{\br \in \Psi_N} X(\br) P[\br | \bx, \Amat, \bpi] \\
  &=& \mu.
\end{eqnarray}
It follows from the strong law of large numbers that $\widehat
\mu_L \rightarrow \mu$ as $L \rightarrow \infty$.

The following finite-sample concentration inequality demonstrates
that the approximation error, $\widehat \mu_N - \mu$, decays
exponentially in the number of importance samples, $L$.

\begin{prop} \label{prop:ISbound1}
Let $\{(X_i, Z_i)\}$ be a sequence of independent and identically
distributed random variables with $X_i \in \{0,1\}$ and $Z_i \in
[0,b]$.  Assume that $\E[Z_i] = 1$ and $\E[Z_i X_i] = \mu$, and
set $\widehat \mu_L = \frac{\sum_{i=1}^L Z_i X_i}{\sum_{i=1}^L
Z_i}$.  Then with probability greater than $1 - \delta$,
\begin{eqnarray}
  \widehat \mu_L - \mu &<& \sqrt{\frac{2 b^2
  \log\frac{2}{\delta}}{L}}.
\end{eqnarray}
\end{prop}

\begin{proof}
From the definitions of $Z_i$ and $X_i$, $Z_i X_i \in [0,b]$.
Applying Hoeffding's inequality \cite{hoeffding63} yields that for
any $t > 0$,
\begin{eqnarray}
  \Pr\left(\sum_{i=1}^L Z_i X_i - L \mu \ge L t\right) &\le&
  e^{-2Lt^2 / b^2},
\end{eqnarray}
and for any $t > 0$,
\begin{eqnarray}
  \Pr\left(\sum_{i=1}^L Z_i - L \le - L t\right) &\le& e^{-2Lt^2 / b^2}.
\end{eqnarray}
Define the event, $E_t = \left\{\sum_{i=1}^L Z_i X_i - L \mu \ge L
t\right\} \bigcup \left\{\sum_{i=1}^L Z_i - L \le - L t\right\}$.
By the union bound, $\Pr(E_t) \le 2e^{-2Lt^2 / b^2}$ for any $t >
0$.  The complement of $E_t$ implies that for $t \in (0,1)$,
\begin{eqnarray}
  \widehat \mu_L - \mu &=& \frac{\sum_{i=1}^L Z_i X_i - L
  \mu}{\sum_{i=1}^L Z_i} + \frac{L \mu}{\sum_{i=1}^L Z_i} - \mu\\
  &<& \frac{L t}{L(1-t)} + \frac{L \mu}{L(1-t)} - \mu\\
  &=& \frac{t(1 + \mu)}{1-t}.
\end{eqnarray}
It follows that $\left\{\widehat \mu_L - \mu \ge
\frac{t(1+\mu)}{1-t}\right\} \subseteq E_t$, and so
$\Pr\left(\widehat \mu_L - \mu \ge \frac{t(1+\mu)}{1-t}\right) \le
\Pr(E_t) \le 2 e^{-2 L t^2 / b^2}$.  Since $\widehat \mu_L \le 1$,
if $\frac{t(1+\mu)}{1-t} + \mu > 1$ then $\Pr\left(\widehat \mu_L
- \mu \ge \frac{t(1+\mu)}{1 - t}\right) = 0$, and the proposition
holds trivially.  Thus, without loss of generality we consider the
case $\frac{t(1+\mu)}{1-t} + \mu \le 1$, or equivalently, $t \le
(1-\mu)/2$.  This restriction on $t$ implies $\frac{t(1+\mu)}{1-t}
\le 2t$, and so we have $\Pr\left(\widehat \mu_L - \mu < 2t\right)
> 1 - 2e^{-2 L t^2 / b^2}$.  Set $\delta = 2e^{-2 L t^2 / b^2}$ to
obtain the desired result.
\end{proof}

We apply Proposition~\ref{prop:ISbound1} to the Monte Carlo
approximations $\{\widehat \alpha_{t',t''}^{(m)}\}$ and
$\{\widehat r_{1,t'}^{(m)}\}$ as follows.  Recall that the Monte
Carlo weights are bounded according to $z_i \in [0,b_m]$, with
$b_m$ as defined in \eqref{eq:mc_bm}.  Define
\begin{eqnarray*}
  B_{\delta', L_m}^m &=& \left(\bigcup_{\genfrac{}{}{0pt}{3}{t',t''=1}{t'\ne t''}}^{N_m} \left\{\widehat
  \alpha_{t',t''}^{(m)} - \bar \alpha_{t',t''}^{(m)} \ge
  \sqrt{\frac{2 b_m^2 \log \frac{2}{\delta'}}{L_m}}
  \right\}\right) \bigcup \left(\bigcup_{t'=1}^{N_m}\left\{\widehat
  r_{1,t'}^{(m)} - \bar r_{1,t'}^{(m)} \ge \sqrt{\frac{2 b_m^2
  \log \frac{2}{\delta'}}{L_m}} \right\}\right).
\end{eqnarray*}
This is a union over $2\binom{N_m}{2} + N_m = N_m^2$ events, each
of which holds with probability at most $\delta'$ according to
Proposition~\ref{prop:ISbound1}.  By the union bound it follows
that $\Pr(B_{\delta',L_m}^m) \le N_m^2 \delta'$. Next, define
\begin{equation}
C_{\delta', L_m}^m = \left\{\sum_{t',t''=1}^{N_m} \left(\widehat
\alpha_{t',t''}^{(m)} - \bar \alpha_{t',t''}^{(m)}\right) +
\sum_{t'=1}^{N_m} \left(\widehat r_{1,t'}^{(m)} - \bar
r_{1,t'}^{(m)}\right) \ge N_m^2 \sqrt{\frac{2 b_m^2 \log
\frac{2}{\delta'}}{L_m}}\right\},
\end{equation}
and observe that $C_{\delta', L_m}^m \subseteq B_{\delta',
L_m}^m$, therefore $\Pr(C_{\delta',L_m}^m) \le
\Pr(B_{\delta',L_m}^m) \le N_m^2 \delta'$. Let $\delta'' = N_m^2
\delta'$ and let $L
> 0$ be a value to be determined later.  For each
$m=1,\dots,T$, set
\begin{eqnarray}
  L_m &=& \frac{2 L N_m^4 b_m^2 \log \frac{2 N_m^2}{\delta''}}{\log
  \frac{1}{\delta''}}, \label{eq:pam_Lm}
\end{eqnarray}
so that
\begin{equation}
  N_m^2 \sqrt{\frac{2 b_m^2 \log \frac{2}{\delta'}}{L_m}}\ =\
  N_m^2 \sqrt{\frac{2 b_m^2 \log \frac{2N_m^2}{\delta''}}{L_m}}\ =\
  \sqrt{\frac{\log \frac{1}{\delta''}}{L}}.
\end{equation}
Then with probability greater than $1 - \delta''$,
\begin{eqnarray}
\sum_{t',t''=1}^{N_m} \left(\widehat \alpha_{t',t''}^{(m)} - \bar
\alpha_{t',t''}^{(m)}\right) + \sum_{t'=1}^{N_m} \left(\widehat
r_{1,t'}^{(m)} - \bar r_{1,t'}^{(m)}\right) &<& \sqrt{\frac{\log
\frac{1}{\delta''}}{L}}. \label{eq:pam_singleObs}
\end{eqnarray}
Recall that $x_{t',i}^{(m)}$ are indicator variables satisfying
$\sum_{i,j=1}^{|S|} x_{t'',i}^{(m)} x_{t',j}^{(m)} = 1$ and
$\sum_{i=1}^{|S|} x_{t',i}^{(m)} = 1$.  Multiplying each term in
\eqref{eq:pam_singleObs} by the appropriate sum of indicators,
rearranging terms, and recalling that importance sample estimates
for different observations are statistically independent, we have
that with probability greater than $(1-\delta'')^T$,
\begin{equation}
\bigcap_{m=1}^T \left\{\sum_{t',t''=1}^{N_m} \sum_{i,j=1}^{|S|}
\left(\widehat \alpha_{t',t''}^{(m)} - \bar
\alpha_{t',t''}^{(m)}\right) x_{t'',i}^{(m)} x_{t',j}^{(m)} +
\sum_{t'=1}^{N_m} \sum_{i=1}^{|S|} \left(\widehat r_{1,t'}^{(m)} -
\bar r_{1,t'}^{(m)}\right) x_{t',i}^{(m)} < \sqrt{\frac{\log
\frac{1}{\delta''}}{L}} \right\},
\end{equation}
which implies that with probability greater than $(1-\delta'')^T$,
\begin{eqnarray}
\sum_{m=1}^T \sum_{t',t''=1}^{N_m} \sum_{i,j=1}^{|S|}
\left(\widehat \alpha_{t',t''}^{(m)} - \bar
\alpha_{t',t''}^{(m)}\right) x_{t'',i}^{(m)} x_{t',j}^{(m)} +
\sum_{m=1}^T \sum_{t'=1}^{N_m} \sum_{i=1}^{|S|} \left(\widehat
r_{1,t'}^{(m)} - \bar r_{1,t'}^{(m)}\right) x_{t',i}^{(m)} &<& T
\sqrt{\frac{\log \frac{1}{\delta''}}{L}}.
\end{eqnarray}
Finally, set $1 - \delta = (1-\delta'')^T$ and multiply through by
$|\log \thetamin| > 0$.  Then with probability greater than $1 -
\delta$,
\begin{eqnarray}
\lefteqn{\sum_{m=1}^T \sum_{t',t''=1}^{N_m} \sum_{i,j=1}^{|S|}
\left(\widehat \alpha_{t',t''}^{(m)} - \bar
\alpha_{t',t''}^{(m)}\right) x_{t'',i}^{(m)} x_{t',j}^{(m)}\ |\log
\thetamin| + \sum_{m=1}^T \sum_{t'=1}^{N_m} \sum_{i=1}^{|S|}
\left(\widehat r_{1,t'}^{(m)} - \bar r_{1,t'}^{(m)}\right)
x_{t',i}^{(m)}\ |\log \thetamin|} \nonumber \\
&<& T\ |\log \thetamin| \sqrt{\frac{-\log\big(1 -
(1-\delta)^{1/T}\big)}{L}}. \hspace{3.25in}
\label{eq:pam_DeltaBound}
\end{eqnarray}

To complete the proof, observe that
\begin{eqnarray}
\widehat \Delta(\widehat \btheta) - \Delta(\widehat
  \btheta)
  &=& \sum_{m=1}^T \sum_{i,j=1}^{|S|} \sum_{t',t''}^{N_m}
  \left(\widehat \alpha_{t',t''}^{(m)} - \bar
  \alpha_{t',t''}^{(m)}\right) x_{t'',i}^{(m)} x_{t',j}^{(m)}
  \left(\log \widehat A_{i,j} - \log A'_{i,j}\right) \nonumber \\
  & & + \sum_{m=1}^T \sum_{i=1}^{|S|} \sum_{t'=1}^{N_m}
  \left(\widehat r_{1,t'}^{(m)} - \bar r_{1,t'}^{(m)}\right)
  x_{t',i}^{(m)} \left(\log \widehat \pi_i - \log \pi'_i\right).
  \label{eq:pam_DeltaBound2}
\end{eqnarray}
By assumption, $\thetamin \le \widehat A_{i,j}, A'_{i,j} \le 1$
for each $(i,j) \in S^2$.  It follows that
\begin{eqnarray}
  \log \widehat A_{i,j} - \log A'_{i,j} &\le& -\log \thetamin
 \ =\ |\log \thetamin|.
\end{eqnarray}
Similarly, $\log \widehat \pi_i - \log \pi'_i \le |\log\thetamin|$
for each $i \in S$.  Apply these bounds in
\eqref{eq:pam_DeltaBound2} to find that the right hand side of
\eqref{eq:pam_DeltaBound2} is no greater than the left hand side
of \eqref{eq:pam_DeltaBound}. Set
\begin{eqnarray}
  \epsilon &=& T\ |\log \thetamin| \sqrt{\frac{\log \frac{1}{1 -
  (1-\delta)^{1/T}}}{L}}. \label{eq:pam_eps}
\end{eqnarray}
Then $\widehat \Delta(\widehat \btheta) - \Delta(\widehat \btheta)
< \epsilon$ with probability greater than $1-\delta$. Solve for
$L$ in \eqref{eq:pam_eps} and plug the resulting value back into
\eqref{eq:pam_Lm} with $\delta'' = 1 - (1-\delta)^{1/T}$ to obtain
the desired result.

\section{Proof of Theorem~\ref{thm:probmono1}}
\label{sec:probmonoproof}

To prove Theorem~\ref{thm:probmono1} we will show that
$\Delta(\widehat \btheta) > (1 - \epsilon) \;\Delta(\btheta^*)$
with high probability, but first we need two preliminary results.
We begin by deriving concentration inequalities for the Monte
Carlo sufficient statistics.  Then we use these bounds to show
that the corresponding M-step parameter estimates, $\widehat
A_{i,j}$ and $\widehat \pi_i$ concentrate about their asymptotic
means, $A^*_{i,j}$ and $\pi^*_{i}$.  From there we construct the
desired bound for $\Delta(\widehat \btheta)$, which implies the
theorem since $\Delta(\btheta^*) \ge 0$ by definition.

The proof of Theorem~\ref{thm:pam} makes use of \emph{additive}
concentration inequalities, bounding the probability of deviations
of the form $\widehat \mu_L - \mu \ge t$.  In this proof we make
use of \emph{relative} concentration inequalities to ensure that
$\widehat \mu_L > (1 + \epsilon) \mu$ with high probability.

\begin{prop} \label{prop:pm1_ISrel}
Let $\{(X_i, Z_i)\}$ be a sequence of independent and identically
distributed random variables with $X_i \in \{0,1\}$ and $Z_i \in
[0,b]$.  Assume that $\E[Z_i] = 1$ and $\E[Z_i X_i] = \mu$, and
set $\widehat \mu_L = \frac{\sum_{i=1}^L Z_i X_i}{\sum_{i=1}^L
Z_i}$, as before.  Then with probability greater than $1 -
\delta$,
\begin{eqnarray*}
  \widehat \mu_L &<& \left(1 + \sqrt{\frac{27 b \log
  \frac{2}{\delta}}{L \mu}}\right)\mu,
\end{eqnarray*}
and with probability greater than $1 - \delta$,
\begin{eqnarray*}
  \widehat \mu_L &>& \left(1 - \sqrt{\frac{27 b \log
  \frac{2}{\delta}}{L \mu}}\right)\mu.
\end{eqnarray*}
\end{prop}

\begin{proof}
Since $X_i \in \{0,1\}$ and $Z_i \in [0,b]$, $Z_i X_i \in [0,b]$
also.  Applying the relative form of Hoeffding's inequality (see,
\emph{e.g.}, Theorem~2.3 in \cite{mcdiarmid98}), we have that for
any $\beta > 0$,
\begin{eqnarray}
  \Pr\left(\sum_{i=1}^L Z_i X_i \ge (1 + \beta) L \mu\right) &\le&
  \exp\left\{\frac{- L \mu \beta^2}{2 b (1 + \beta /3)}\right\}.
\end{eqnarray}
If $\beta \le 1$ then $2(1 + \beta/3) < 3$, and so for $\beta \in
(0,1]$,
\begin{eqnarray}
  \Pr\left(\sum_{i=1}^L Z_i X_i \ge (1 + \beta) L \mu\right) &\le&
  \exp\left\{\frac{- L \mu \beta^2}{3 b}\right\},
  \label{eq:pm_hoeff1}
\end{eqnarray}
which suffices for our application.  Also, for any $\gamma > 0$,
\begin{eqnarray}
  \Pr\left(\sum_{i=1}^L Z_i \le (1 - \gamma) L \right) &\le&
  \exp\left\{\frac{- L \gamma^2}{2 b}\right\}. \label{eq:pm_hoeff2}
\end{eqnarray}
Suppose the events
\begin{eqnarray}
\left\{\sum_{i=1}^L Z_i X_i < (1 + \beta) L \mu\right\} & \mbox{
and } & \left\{\sum_{i=1}^L Z_i > (1 - \gamma) L\right\}
\label{eq:pm_events1}
\end{eqnarray}
occur simultaneously.  Then for $0 < \gamma < 1$,
\begin{eqnarray}
  \widehat \mu_L &>& \left(\frac{1 + \beta}{1 - \gamma}\right) \mu.
\end{eqnarray}
Since we will apply the union bound, we balance the exponential
rates in \eqref{eq:pm_hoeff1} and \eqref{eq:pm_hoeff2} by setting
$\gamma = \beta \sqrt{\frac{2}{3}\mu} < 1$.  Solving
\begin{equation}
  \frac{1 + \beta}{1 - \gamma} = \frac{1 + \beta}{1 -
  \beta\sqrt{\frac{2}{3}\mu}} = 1 + \epsilon
\end{equation}
for $\beta$ in terms of $\epsilon$ leads to
\begin{eqnarray}
  \beta &=& \frac{\epsilon}{1 + \sqrt{\frac{2}{3}\mu} + \epsilon
  \sqrt{\frac{2}{3}\mu}}. \label{eq:pm_beta1}
\end{eqnarray}
In order to ensure that $\beta \le 1$ we restrict
\begin{eqnarray}
  \epsilon &\le& \frac{1 + 1}{1 - \sqrt{\frac{2}{3}\mu}}
  - 1\\
  &=& \frac{1 + \sqrt{\frac{2}{3}\mu}}{1 - \sqrt{\frac{2}{3}\mu}}.
\end{eqnarray}
Note that the right hand side of the expression above is at least
1 for all $\mu \in [0,1]$.  Apply the union bound with the
complements of the events in \eqref{eq:pm_events1} using
\eqref{eq:pm_beta1} in the exponent, and observe that $1 +
\sqrt{\frac{2}{3}\mu} + \epsilon\sqrt{\frac{2}{3}\mu} \le 3$ for
all $\mu \in [0,1]$ and $\epsilon \in (0,1)$ to find that
$\Pr\left(\widehat \mu_L \le (1 + \epsilon) \mu\right) \le 2
e^{-L\mu\epsilon^2 / 27 b^2}$.  Set $\delta = 2 e^{-L\mu\epsilon^2
/ 27 b^2}$ to obtain the first claim.  The proof of the second
claim follows a similar sequence of steps. See
\cite{RabbatFigueiredoNowak_TR} for the full details.
\end{proof}

Application of the union bound yields the following.

\begin{cor} \label{cor:pm1_ISrel}
With probability greater than $1 - \delta$,
\begin{equation}
\left(1 - \sqrt{\frac{27 b \log \frac{4}{\delta}}{L \mu}}\right)
\mu\ <\ \widehat \mu_L\ <\ \left(1 + \sqrt{\frac{27 b \log
\frac{4}{\delta}}{L \mu}}\right)\mu. \label{eq:ISrelTwoWay}
\end{equation}
\end{cor}

Next we apply Corollary~\ref{cor:pm1_ISrel} to the Monte Carlo
approximations, $\{\widehat r_{1,t'}^{(m)}\}$ and $\{\widehat
\alpha_{t',t''}^{(m)}\}$ towards showing that $\widehat A_{i,j}$
and $\widehat \pi_i$ do not deviate too greatly from $A_{i,j}^*$
and $\pi_i^*$.  Recall the exact M-step expressions for $\pi^*_i$
and $A^*_{i,j}$ given by \eqref{eqn:Mstep_A}.  The corresponding
expressions for $\widehat \pi_i$ and $\widehat A_{i,j}$ are found
by replacing each $\bar r_{1,t'}^{(m)}$ and $\bar
\alpha_{t',t''}^{(m)}$ with $\widehat r_{1,t'}^{(m)}$ and
$\widehat \alpha_{t',t''}^{(m)}$. By appropriately bounding the
numerators and denominators of $\widehat A_{i,j}$ and $\widehat
\pi_i$ we obtain the following result.

\begin{prop} \label{prop:pm_params}
Let $L > 0$ and $\delta > 0$ be given.  Assume that there exists
$\lambda > 0$ such that $\bar r_{1,t'}^{(m)} \ge \lambda$ and
$\bar \alpha_{t',t''}^{(m)} \ge \lambda$ for all $m=1,\dots,T$ and
$t', t''=1,\dots,N_m$. If at least $L_m \ge \frac{27 b_m
L}{\lambda}$ importance samples are used, then with probability at
least $1 - (\sum_{m=1}^T N_m^2) \delta$,
\begin{eqnarray}
\left(\bigcap_{i,j=1}^{|S|} \left\{\widehat A_{i,j} >
\left(\frac{1 - \sqrt{\frac{\log \frac{4}{\delta}}{L}}}{1 +
\sqrt{\frac{\log \frac{4}{\delta}}{L}}}\right)
A^*_{i,j}\right\}\right) \bigcap \left(\bigcap_{i=1}^{|S|}
\left\{\widehat \pi_i > \left(\frac{1 - \sqrt{\frac{\log
\frac{4}{\delta}}{L}}}{1 + \sqrt{\frac{\log
\frac{4}{\delta}}{L}}}\right) \pi^*_i \right\}\right).
\end{eqnarray}
\end{prop}

\begin{proof}
First recall that there are $2\binom{N_m}{2} + N_m = N_m^2$
sufficient statistics associated with the $m$th observation: one
$\alpha_{t',t''}^{(m)}$ for each of the $2\binom{N_m}{2}$ possible
transitions and one $r_{1,t'}^{(m)}$ for each possible initial
state. Then, in total there are $\sum_{m=1}^T N_m^2$ sufficient
statistics to calculate in the E-step.  Applying the union bound
in conjunction with \eqref{eq:ISrelTwoWay} we have that with
probability greater than $1 - (\sum_{m=1}^T N_m^2)\delta$,
\begin{equation}
\bar \alpha_{t',t''}^{(m)} - \sqrt{\frac{27 b_m \bar
\alpha_{t',t''}^{(m)} \log \frac{4}{\delta}}{L_m}}\ <\ \widehat
\alpha_{t',t''}^{(m)}\ <\ \bar \alpha_{t',t''}^{(m)} +
\sqrt{\frac{27 b_m \bar \alpha_{t',t''}^{(m)} \log
\frac{4}{\delta}}{L_m}},
\end{equation}
for all $m=1,\dots,T$ and $t',t''=1,\dots,N_m$, and
\begin{equation}
\bar r_{1,t'}^{(m)} - \sqrt{\frac{27 b_m \bar r_{1,t'}^{(m)} \log
\frac{4}{\delta}}{L_m}}\ <\ \widehat r_{1,t'}^{(m)}\ <\ \bar
r_{1,t'}^{(m)} + \sqrt{\frac{27 b_m \bar r_{1,t'}^{(m)} \log
\frac{4}{\delta}}{L_m}},
\end{equation}
for all $m=1,\dots,T$ and $t'=1,\dots,N_m$.  Based on the
assumption that $\bar \alpha_{t',t''}^{(m)} \ge \lambda$ and $\bar
r_{1,t'}^{(m)} \ge \lambda$, taking $L_m \ge 27 b_m L / \lambda$
guarantees that
\begin{eqnarray}
L_m &\ge& \max\left(\max_{t',t''=1,\dots,N_m} \frac{27 b_m L}{\bar
\alpha_{t',t''}^{(m)}};\ \max_{t'=1,\dots,N_m} \frac{27 b_m
L}{\bar r_{1,t'}^{(m)}} \right).
\end{eqnarray}
Then with probability greater than $1 - (\sum_{m=1}^T
N_m^2)\delta$,
\begin{equation}
\left(1 - \sqrt{\frac{\log \frac{4}{\delta}}{L}}\right) \bar
\alpha_{t',t''}^{(m)}\ <\ \widehat \alpha_{t',t''}^{(m)}\ <\
\left(1 + \sqrt{\frac{\log \frac{4}{\delta}}{L}}\right) \bar
\alpha_{t',t''}^{(m)}, \label{eq:pm_alphaTwoWay}
\end{equation}
for all $m=1,\dots,T$ and $t',t''=1,\dots,N_m$, and
\begin{equation}
\left(1 - \sqrt{\frac{\log \frac{4}{\delta}}{L}}\right) \bar
r_{1,t'}^{(m)}\ <\ \widehat r_{1,t'}^{(m)}\ <\ \left(1 +
\sqrt{\frac{\log \frac{4}{\delta}}{L}}\right) \bar r_{1,t'}^{(m)},
\label{eq:pm_rTwoWay}
\end{equation}
for all $m=1,\dots,T$ and $t'=1,\dots,N_m$.

Equation \eqref{eq:pm_alphaTwoWay} implies that for each $(i,j)
\in S^2$,
\begin{eqnarray*}
\sum_{m=1}^T \sum_{t',t''=1}^{N_m} \widehat \alpha_{t',t''}^{(m)}
x_{t'',i}^{(m)} x_{t',j}^{(m)} &>& \left(1 - \sqrt{\frac{\log
\frac{4}{\delta}}{L}}\right) \sum_{m=1}^T \sum_{t',t''=1}^{N_m}
\bar \alpha_{t',t''}^{(m)} x_{t'',i}^{(m)} x_{t',j}^{(m)},
\end{eqnarray*}
and for each $i \in S$,
\begin{eqnarray*}
\sum_{k=1}^{|S|} \sum_{m=1}^T \sum_{t',t''}^{(m)} \widehat
\alpha_{t',t''}^{(m)} x_{t'',i}^{(m)} x_{t',k}^{(m)} &<& \left(1 +
\sqrt{\frac{\log \frac{4}{\delta}}{L}}\right) \sum_{k=1}^{|S|}
\sum_{m=1}^T \sum_{t',t''=1}^{N_m} \bar \alpha_{t',t''}^{(m)}
x_{t'',i}^{(m)} x_{t',k}^{(m)}.
\end{eqnarray*}
Taking the ratio of these two expressions yields the desired
result for $\widehat A_{i,j}$ and $A^*_{i,j}$.

Similarly, \eqref{eq:pm_rTwoWay} implies that for each $i$,
\begin{eqnarray}
\sum_{m=1}^T \sum_{t'=1}^{N_m} \widehat r_{1,t'}^{(m)}
x_{t',i}^{(m)} &>& \left(1 - \sqrt{\frac{\log
\frac{4}{\delta}}{L}}\right) \sum_{m=1}^T \sum_{t'}^{N_m} \bar
r_{1,t'}^{(m)} x_{t',i}^{(m)},
\end{eqnarray}
and for each $i$,
\begin{eqnarray}
\sum_{m=1}^T \sum_{t'=1}^{N_m} \widehat r_{1,t'}^{(m)}
x_{t',i}^{(m)} &<& \left(1 + \sqrt{\frac{\log
\frac{4}{\delta}}{L}}\right) \sum_{m=1}^T \sum_{t'}^{N_m} \bar
r_{1,t'}^{(m)} x_{t',i}^{(m)}.
\end{eqnarray}
Taking the ratio of these two expressions yields the desired
result for $\widehat \pi_i$ and $\pi^*_i$.
\end{proof}

The remainder of the proof of Theorem~\ref{thm:probmono1} is now
fairly straightforward.  Let $\delta > 0$ be the value given in
the statement of Theorem~\ref{thm:probmono1}.  Monotonicity of the
logarithm in conjunction with Proposition~\ref{prop:pm_params}
implies that with probability greater than $1 - \delta$, for all
$i,j \in S$,
\begin{eqnarray}
  \log \widehat A_{i,j} &>& \log A^*_{i,j} + \log\left(
\frac{1 - \sqrt{\frac{\log 4 \sum_{m=1}^T N_m^2 - \log
\delta}{L}}}{1 + \sqrt{\frac{\log 4 \sum_{m=1}^T N_m^2 - \log
\delta}{L}}}\right), \\
\log \widehat \pi_i &>& \log \pi^*_i + \log\left(\frac{1 -
\sqrt{\frac{\log 4 \sum_{m=1}^T N_m^2 - \log \delta}{L}}}{1 +
\sqrt{\frac{\log 4 \sum_{m=1}^T N_m^2 - \log \delta}{L}}}\right).
\end{eqnarray}
Multiply through by either $\sum_{m=1}^T \sum_{t',t''=1}^{N_m}
\bar \alpha_{t',t''}^{(m)} x_{t'',i}^{(m)} x_{t',j}^{(m)} > 0$ or
$\sum_{m=1}^T \sum_{t'=1}^{N_m} \bar \alpha_{t',t''}^{(m)}
x_{t',i}^{(m)} > 0$ as appropriate, and sum over $i$ and $j$ to
obtain
\begin{equation}
\begin{array}{l}
Q(\widehat \btheta; \btheta') \ > \ Q(\btheta^*; \btheta') \\
\hspace{0.4in} + \left(\displaystyle \sum_{i,j=1}^{|S|}
\sum_{m=1}^T \sum_{t',t''=1}^{N_m} \bar \alpha_{t',t''}^{(m)}
x_{t'',i}^{(m)} x_{t',j}^{(m)} + \sum_{i=1}^{|S|} \sum_{m=1}^T
\sum_{t'=1}^{N_m} \bar r_{1,t'}^{(m)} x_{t',i}^{(m)}\right)
\log\left(\frac{1 - \sqrt{\frac{\log 4 \sum_{m=1}^T N_m^2 - \log
\delta}{L}}}{1 + \sqrt{\frac{\log 4 \sum_{m=1}^T N_m^2 - \log
\delta}{L}}}\right).
\end{array} \label{eq:pm_QBound1}
\end{equation}
By the definitions of $x_{t',i}^{(m)}$, $\bar r_{1,t'}^{(m)}$, and
$\bar \alpha_{t',t''}^{(m)}$ given in Section~\ref{sec:em} we have
$\sum_{t'=1}^{N_m} \bar r_{1,t'}^{(m)} = 1$,
$\sum_{t',t''=1}^{N_m} \bar \alpha_{t',t''}^{(m)} = N_m -1$,
$\sum_{i,j=1}^{|S|} x_{t'',i}^{(m)} x_{t',j}^{(m)} = 1$, and
$\sum_{i=1}^{|S|} x_{t',i}^{(m)}=1$.  It follows that
\begin{eqnarray}
\sum_{i,j=1}^{|S|} \sum_{m=1}^T \sum_{t',t''=1}^{N_m} \bar
\alpha_{t',t''}^{(m)} x_{t'',i}^{(m)} x_{t',j}^{(m)} +
\sum_{i=1}^{|S|} \sum_{m=1}^T \sum_{t'=1}^{N_m} \bar
r_{1,t'}^{(m)} x_{t',i}^{(m)} &=& \sum_{m=1}^T N_m.
\end{eqnarray}
Subtract $Q(\btheta'; \btheta')$ from both sides of
\eqref{eq:pm_QBound1} to obtain that with probability greater than
$1 - \delta$,
\begin{eqnarray}
\Delta(\widehat \btheta) &>& \Delta(\btheta^*) +
\left(\sum_{m=1}^T N_m\right) \log\left(\frac{1 - \sqrt{\frac{\log
4 \sum_{m=1}^T N_m^2 - \log \delta}{L}}}{1 + \sqrt{\frac{\log 4
\sum_{m=1}^T N_m^2 - \log \delta}{L}}}\right).
\end{eqnarray}
Let $\epsilon > 0$ be the value given in the statement of
Theorem~\ref{thm:probmono1} and set
\begin{eqnarray}
\left(\sum_{m=1}^T N_m\right) \log\left(\frac{1 - \sqrt{\frac{\log
4 \sum_{m=1}^T N_m^2 - \log \delta}{L}}}{1 + \sqrt{\frac{\log 4
\sum_{m=1}^T N_m^2 - \log \delta}{L}}}\right) &=& -\epsilon
\Delta(\btheta^*).
\end{eqnarray}
Solving for $L$ yields
\begin{eqnarray}
L &=& \left(\frac{1 + \exp\left\{\frac{-\epsilon
\Delta(\btheta^*)}{\sum_{m=1}^T N_m}\right\}}{1 -
\exp\left\{\frac{-\epsilon \Delta(\btheta^*)}{\sum_{m=1}^T
N_m}\right\}}\right)^2 \log\left(\frac{4\sum_{m=1}^T
N_m^2}{\delta}\right). \label{eq:pm_L}
\end{eqnarray}

Recall the well known inequality: $u \ge \log(1 + u)$ for $u \ge
0$.  Putting $u = \frac{\epsilon \Delta(\btheta^*)}{\sum_{m=1}^T
N_m} \ge 0$ leads to
\begin{eqnarray}
\frac{\epsilon \Delta(\btheta^*)}{\sum_{m=1}^T N_m} &\ge& 1 +
\frac{\epsilon \Delta(\btheta^*)}{\sum_{m=1}^T N_m}.
\end{eqnarray}
Take the exponential, which is a monotonic transformation, and
then invert the resulting expression to obtain
\begin{eqnarray}
\exp\left\{\frac{-\epsilon\Delta(\btheta^*)}{\sum_{m=1}^T
N_m}\right\} &\le& \left(1 + \frac{\epsilon
\Delta(\btheta^*)}{\sum_{m=1}^T N_m}\right)^{-1}.
\end{eqnarray}
It follows that
\begin{eqnarray}
  \frac{1 + \exp\left\{\frac{-\epsilon\Delta(\btheta^*)}{\sum_{m=1}^T
N_m}\right\}}{1 -
\exp\left\{\frac{-\epsilon\Delta(\btheta^*)}{\sum_{m=1}^T
N_m}\right\}} &\le& \frac{2 \sum_{m=1}^T N_m +
\epsilon\Delta(\btheta^*)}{\epsilon \Delta(\btheta^*)}.
\end{eqnarray}
Using this last result in \eqref{eq:pm_L}, together with the
choice of $L_m$ from Proposition~\ref{prop:pm_params}, we find
that if we use
\begin{eqnarray}
L_m &=& \frac{27 b_m}{\lambda} \left(\frac{2 \sum_{m=1}^T N_m +
\epsilon\Delta(\btheta^*)}{\epsilon \Delta(\btheta^*)}\right)^2
\log\left(\frac{4 \sum_{m=1}^T N_m^2}{\delta}\right)
\end{eqnarray}
importance samples for the $m$th observation in the Monte Carlo
E-step, then $\Delta(\widehat \btheta) \ge (1 -
\epsilon)\Delta(\btheta^*)$ with probability greater than $1 -
\delta$. Since $\Delta(\btheta^*) \ge 0$ by definition we may take
$\epsilon = 1$.  Then $\Delta(\widehat \btheta^*) \ge 0$ with
probability greater than $1 - \delta$.

\bibliographystyle{plain}
\bibliography{netStructure}

\begin{thebibliography}{10}

\bibitem{brainConnectivity05}
{\em International Workshop on Brain Connectivity}, 2005.
\newblock \url{http://www.ccs.fau.edu/~bc2005/welcome.html}.

\bibitem{bernardo94}
J.~Bernardo and A.~Smith.
\newblock {\em Bayesian Theory}.
\newblock John Wiley \& Sons, 1994.

\bibitem{booth01}
J.~G. Booth, J.~P. Hobert, and W.~S. Jank.
\newblock A survey of {Monte} {Carlo} algorithms for maximizing the likelihood
  of a two-stage hierarchical model.
\newblock {\em Statistical Modelling}, 1:333--349, 2001.

\bibitem{boyles83}
R.~A. Boyles.
\newblock On the convergence of the {EM} algorithm.
\newblock {\em Journal of the Royal Statistical Society B}, 45(1):47--50, 1983.

\bibitem{caffo05}
B.~S. Caffo, W.~Jank, and G.~L. Jones.
\newblock Ascent-based {M}onte {C}arlo {EM}.
\newblock {\em Journal of the Royal Statistical Society B}, 67(2):235--252,
  2005.

\bibitem{coates02}
M.~Coates, A.~O. Hero, R.~Nowak, and B.~Yu.
\newblock Internet tomography.
\newblock {\em IEEE Signal Processing Magazine}, 19(3):47--65, 2002.

\bibitem{figueiredo02}
M.~A.~T. Figueiredo and A.~K. Jain.
\newblock Unsupervised learning of finite mixture models.
\newblock {\em IEEE Transactions on Pattern Analysis and Machine Intelligence},
  24(3):381--396, March 2002.

\bibitem{koller03}
N.~Friedman and D.~Koller.
\newblock Being {Bayesian} about {Bayesian} network structure: A {Bayesian}
  approach to structure discovery in {Bayesian} networks.
\newblock {\em Machine Learning}, 50(1--2):95--125, 2003.

\bibitem{heckerman95}
D.~Heckerman, D.~Geiger, and D.~Chickering.
\newblock Learning {Bayesian} networks: The combination of knowledge and
  statistical data.
\newblock {\em Machine Learning}, 20:197--243, 1995.

\bibitem{hoeffding63}
W.~J. Hoeffding.
\newblock Probability inequalities for sums of bounded random variables.
\newblock {\em Journal of the {A}merican Statistical Association}, 58:713--721,
  1963.

\bibitem{jank05}
W.~Jank.
\newblock Stochastic variants of the {EM} algorithm: {Monte} {Carlo},
  quasi-{Monte} {Carlo} and more.
\newblock In {\em Proc. of the American Statistical Association}, Minneapolis,
  Minnesota, August 2005.

\bibitem{justice05}
D.~Justice and A.~O. Hero.
\newblock Estimation of message source and destination from link intercepts.
\newblock Submitted to \textit{IEEE Trans. on Information Forensics and
  Security}, April 2005.

\bibitem{klipp05}
E.~Klipp, R.~Herwig, A.~Kowald, C.~Wierling, and H.~Lehrach.
\newblock {\em Systems Biology in Practice: Concepts, Implementation and
  Application}.
\newblock John Wiley and Sons, 2005.

\bibitem{kubica03}
J.~Kubica, A.~Moore, D.~Cohn, and J.~Schneider.
\newblock {cGraph}: A fast graph-based method for link analysis and queries.
\newblock In {\em Proc. IJCAI Text-Mining and Link-Analysis Workshop},
  Acapulco, Mexico, August 2003.

\bibitem{liu01}
J.~S. Liu.
\newblock {\em Monte Carlo Strategies in Scientific Computing}.
\newblock Springer, 2001.

\bibitem{liu04}
Y.~Liu and H.~Zhao.
\newblock A computational approach for ordering signal transduction pathway
  components from genomics and proteomics data.
\newblock {\em BMC Bioinformatics}, 5(158), October 2004.

\bibitem{mcdiarmid98}
C.~Mc{D}iarmid.
\newblock Concentration.
\newblock In M.~Habib, C.~Mc{D}iarmid, J.~Ramirez-Alfonsin, and B.~Reed,
  editors, {\em Probabilistic Methods for Algorithmic Discrete Mathematics},
  pages 195--248. Springer-Verlag, New York, 1998.

\bibitem{newman06}
M.~Newman, A.~L. Barabasi, and D.~J. Watts.
\newblock {\em The Structure and Dynamics of Networks}.
\newblock Princeton University Pres, 2006.

\bibitem{palsson06}
B.~O. Palsson.
\newblock {\em Systems Biology: Properties of Reconstructed Networks}.
\newblock Cambridge University Press, 2006.

\bibitem{RabbatFigueiredoNowak_TR}
M.~G. Rabbat, M.~A.~T. Figueiredo, and Robert~D. Nowak.
\newblock Network inference from co-occurrences.
\newblock Technical report ECE-06-02, Department of Electrical and Computer
  Engineering, University of Wisconsin-Madison, April 2006.

\bibitem{rabbat05}
M.~G. Rabbat, J.~R. Treichler, S.~L. Wood, and M.~G. Larimore.
\newblock Understanding the topology of a telephone network via
  internally-sensed network tomography.
\newblock In {\em Proc. IEEE International Confernece on Acoustics, Speech, and
  Signal Processing}, volume~3, pages 977--980, Philadelphia, PA, March 2005.

\bibitem{RobertCasella}
C.~Robert and G.~Casella.
\newblock {\em Monte Carlo Statistical Methods}.
\newblock Springer Verlag, New York, 1999.

\bibitem{sporns04}
O.~Sporns, D.~R. Chialvo, M.~Kaiser, and C.~C. Hilgetag.
\newblock Organization, development and function of complex brain networks.
\newblock {\em Trends in Cognitive Science}, 8(9), 2004.

\bibitem{sporns02}
O.~Sporns and G.~Tononi.
\newblock Classes of network connectivity and dynamics.
\newblock {\em Complexity}, 7(1):28--38, 2002.

\bibitem{wasserman94}
S.~Wasserman, K.~Faust, D.~Iacobucci, and M.~Granovetter.
\newblock {\em Social Network Analysis: Methods and Applications}.
\newblock Cambridge University Press, 1994.

\bibitem{wei90}
G.~C.~G. Wei and M.~A. Tanner.
\newblock A {Monte} {Carlo} implementation of the {EM} algorithm and the poor
  man's data augmentation algorithms.
\newblock {\em Journal of the American Statistical Association}, 85:699--704,
  1990.

\bibitem{wu83}
C.~F.~J. Wu.
\newblock On the convergence properties of the {EM} algorithm.
\newblock {\em Annals of Statistics}, 11(1):95--103, 1983.

\bibitem{zhu05}
D.~Zhu, A.~O. Hero, H.~Cheng, R.~Khanna, and A.~Swaroop.
\newblock Network constrained clustering for gene microarray data.
\newblock {\em Bioinformatics}, 21(21):4014--4020, 2005.

\end{thebibliography}

\end{document}